\documentclass[final]{elsarticle}

\usepackage{hyperref,color,amsmath,amssymb,amsfonts,comment,ulem}

\journal{Neural Networks (Published, DOI: 10.1016/j.neunet.2019.03.005)}

\biboptions{authoryear}

\begin{document}

\begin{frontmatter}

\title{Recent Advances in Physical Reservoir Computing: \\ A Review}

\author[]{Gouhei~Tanaka$^{a,b,}$\footnote[1]{Correspondence and requests for materials should be addressed to G.T. (gouhei@sat.t.u-tokyo.ac.jp)},
Toshiyuki Yamane$^c$, Jean Benoit H{\'e}roux$^c$, \\ Ryosho Nakane$^{a,b}$, Naoki Kanazawa$^c$, Seiji Takeda$^c$, \\ Hidetoshi Numata$^c$, Daiju Nakano$^c$, and Akira Hirose$^{a,b}$}

\address{$^a$Institute for Innovation in International Engineering Education, Graduate School of Engineering, The University of Tokyo, Tokyo 113-8656, Japan\\
$^b$Department of Electrical Engineering and Information Systems, Graduate School of Engineering, The University of Tokyo, Tokyo 113-8656, Japan\\
$^c$IBM Research -- Tokyo, Kanagawa 212-0032, Japan
}





\begin{abstract}
Reservoir computing is a computational framework suited for temporal/sequential data processing. It is derived from several recurrent neural network models, including echo state networks and liquid state machines. A reservoir computing system consists of a reservoir for mapping inputs into a high-dimensional space and a readout for pattern analysis from the high-dimensional states in the reservoir. The reservoir is fixed and only the readout is trained with a simple method such as linear regression and classification. Thus, the major advantage of reservoir computing compared to other recurrent neural networks is fast learning, resulting in low training cost. Another advantage is that the reservoir without adaptive updating is amenable to hardware implementation using a variety of physical systems, substrates, and devices. In fact, such physical reservoir computing has attracted increasing attention in diverse fields of research. The purpose of this review is to provide an overview of recent advances in physical reservoir computing by classifying them according to the type of the reservoir. We discuss the current issues and perspectives related to physical reservoir computing, in order to further expand its practical applications and develop next-generation machine learning systems.
\end{abstract}

\begin{keyword}
Neural networks, machine learning, reservoir computing, nonlinear dynamical systems, neuromorphic device
\MSC[2010] 68Txx\sep 37N20
\end{keyword}

\end{frontmatter}


\clearpage
\tableofcontents

\clearpage
\begin{table}[htb]
\begin{center}
{\bf Abbreviations in alphabetical order}\\
\begin{tabular}{|ll|}
\hline
AD: & Analog-to-Digital\\
ANN: & Artificial Neural Network \\
ASIC: & Application Specific Integrated Circuit\\
ASN: & Atomic Switch Network\\
BMI: & Brain Machine Interface\\
BPDC: & Backpropagation Decorrelation\\
BPTT: & Backpropagation Through Time\\
CA: & Cellular Automaton\\
CPU: & Central Processing Unit\\
DA: & Digital-to-Analog\\
DNA: & Deoxyribonucleic acid\\
ECG: & Electrocardiogram\\
EEG: & Electroencephalogram \\
EMG: & Electromyogram \\
ESN: & Echo State Network\\
fMRI: & functional Magnetic Resonance Imaging\\
FNN: & Feedforward Neural Network \\
FORCE: & First Order Reduced and Controlled Error\\
FPGA: & Field-Programmable Gate Array\\
GPU: & Graphics Processing Unit\\
LIF: & Leaky Integrate-and-Fire\\
LLG: & Landau-Lifshitz-Gilbert\\
LSM: & Liquid State Machine\\
LSTM: & Long Short-Term Memory\\
MEA: & Microelectrode Array\\
MFCC: & Mel-Frequency Cepstrum Coefficient\\
MTJ: & Magnetic Tunnel Junction\\
NARMA: & Nonlinear Autoregressive Moving Average\\
ODE: & Ordinary Differential Equation\\
PM & Particulate Matter \\
RC: & Reservoir Computing\\
RNN: & Recurrent Neural Network\\
RTRL: & Real-Time Recurrent Learning\\
SNN: & Spiking Neural Networks\\
SOA: & Semiconductor Optical Amplifier\\
STDP: & Spike-Timing-Dependent Plasticity\\
STO: & Spin Torque Oscillator\\
VCSEL: & Vertical Cavity Surface Emitting Laser\\
VLSI: & Very Large Scale Integration\\
XOR: & Exclusive OR \\
YIG: & Yttrium Iron Garnet\\
\hline
\end{tabular}
\end{center}
\end{table}

\clearpage

\section{Introduction \label{sec:intro}}

Artificial neural networks (ANNs) constitute the core information processing technology in the fields of artificial intelligence and machine learning, which have witnessed remarkable progress in recent years, and they are expected to be increasingly employed in real-world applications \citep{samarasinghe2016neural}. ANNs are computational models that mimic biological neural networks. They are represented by a network of neuron-like processing units interconnected via synapse-like weighted links. Network architectures of ANNs are typically classified into feedforward networks \citep{schmidhuber2015deep} and recurrent networks \citep{mandic2001recurrent}, the choice of which depends on the type of computational task. Feedforward neural networks (FNNs) are mainly used for {\it static} (non-temporal) data processing, as individual input data are independently processed even if they are given sequentially. In short, FNNs are capable of approximating nonlinear input-output functions. On the other hand, recurrent neural networks (RNNs) are suited for {\it dynamic} (temporal) data processing, as they can embed temporal dependence of the inputs into their dynamical behavior. In other words, RNNs are capable of representing dynamical systems driven by sequential inputs owing to their feedback connections.

Reservoir computing (RC) is originally an RNN-based framework and is therefore suitable for temporal/sequential information processing \citep{jaeger2004harnessing}. Specifically, RC is a unified computational framework \citep{verstraeten2007experimental,lukovsevivcius2009reservoir}, derived from independently proposed RNN models, such as echo state networks (ESNs) \citep{jaeger2001echo} and liquid state machines (LSMs) \citep{maass2002real}. The backpropagation decorrelation (BPDC) learning rule \citep{steil2004backpropagation,steil2007online} for RNNs is also regarded as a predecessor of RC. Similar concepts and models in special cases were reported in earlier studies as summarized in \citep{Jaeger:2007}, including sequential associative memory models \citep{gallant1988experiments}, neural oscillator network models for learning handwriting movements \citep{schomaker1991simulation,schomaker1992neural}, context reverberation networks consisting of linear threshold units for sequential learning \citep{kirby1990neurodynamics,kirby1991context}, cortico-striatal models for context-dependent sequence learning \citep{dominey1995model,dominey1995complex}, and biological neural network models for temporal pattern discrimination \citep{buonomano1995temporal}.

\begin{figure}[t]
\begin{center}
\includegraphics[width=1.0\hsize]{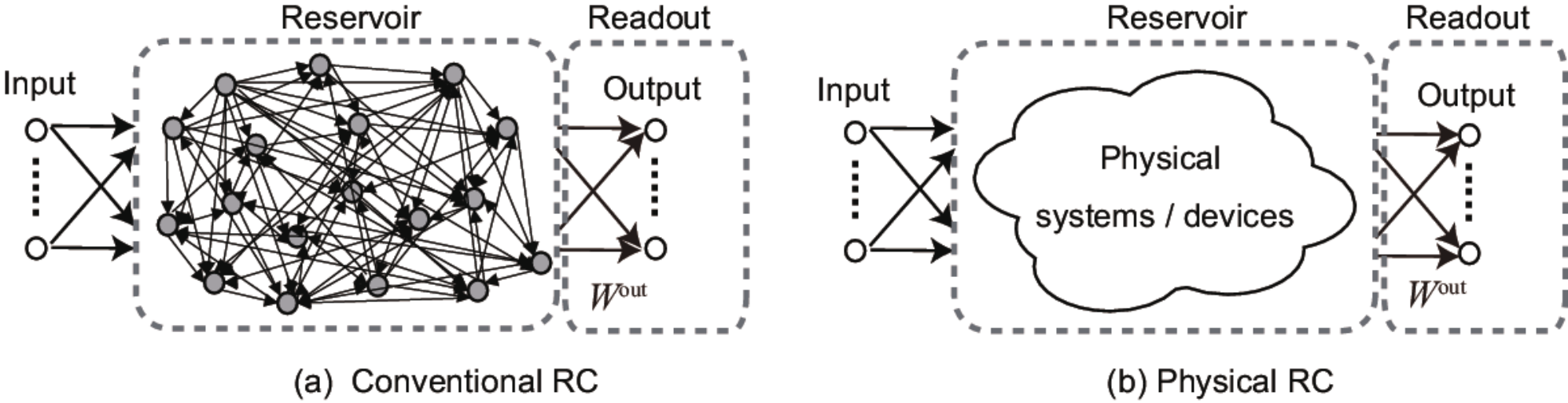}
\caption{RC frameworks where the reservoir is fixed and only the readout weights $W^{\rm out}$ are trained. (a) A conventional RC system with an RNN-based reservoir as in ESNs and LSMs. (b) A physical RC system in which the reservoir is realized using a physical system or device.}
\label{fig:rc}
\end{center}
\end{figure}

In RC, input data are transformed into spatiotemporal patterns in a high-dimensional space by an RNN in the {\it reservoir}. Then, a pattern analysis from the spatiotemporal patterns is performed in the {\it readout}, as shown in Fig.~\ref{fig:rc}(a). The main characteristic of RC is that the input weights ($W^{\rm in}$) and the weights of the recurrent connections within the reservoir ($W$) are not trained whereas only the readout weights ($W^{\rm out}$) are trained with a simple learning algorithm such as linear regression. This simple and fast training process makes it possible to drastically reduce the computational cost of learning compared with standard RNNs, which is the major advantage of RC \citep{jaeger2002tutorial}. RC models have been successfully applied to many computational problems, such as temporal pattern classification, prediction, and generation. To enhance computational performance of RC, it is necessary to appropriately represent sample data and optimally design the RNN-based reservoir. The methods for obtaining effective reservoirs, which have been summarized in \citep{lukovsevivcius2009reservoir}, are categorized into task-independent generic guidelines and task-dependent reservoir adjustments.

The role of the reservoir in RC is to nonlinearly transform sequential inputs into a high-dimensional space such that the features of the inputs can be efficiently read out by a simple learning algorithm. Therefore, instead of RNNs, other nonlinear dynamical systems can be used as reservoirs. In particular, {\it physical RC} using reservoirs based on physical phenomena has recently attracted increasing interest in many research areas (Fig.~\ref{fig:rc}(b)). Various physical systems, substrates, and devices have been proposed for realizing RC. A motivation for physical implementation of reservoirs is to realize fast information processing devices with low learning cost. For hardware implementation of normal RNNs where training is necessary, we often rely on advanced technologies of neural network hardware \citep{misra2010artificial} and neuromorphic hardware \citep{hasler2013finding}. In contrast, physical implementation of reservoirs can be achieved using a variety of physical phenomena in the real world, because a mechanism for adaptive changes for training is not necessary. Actually, physical RC is one of the candidates of unconventional computing paradigms based on novel hardware \citep{hadaeghi2017unconventional}. Although design principles for conventional RC, such as ESNs \citep{ozturk2007analysis,lukovsevivcius2012practical} and LSMs \citep{maass2011liquid}, have been examined comprehensively, the following issues require further investigation: how to design physical reservoirs for achieving high computational performance and how much computational power can be attained by individual physical RC systems.

The purpose of this review is to provide an overview of recent advances in RC, with a special focus on physical RC. Published studies on physical RC can be found not only in neural network journals but also in specialized journals related to the respective physical systems. Our objective is to provide a comprehensive viewpoint with regard to interdisciplinary studies on physical RC by classifying them according to the type of the physical phenomenon utilized for the reservoir. Toward this end, we summarize the characteristics of individual physical reservoirs. Our classification, which highlights the similarities and differences among different physical reservoirs, is useful for gaining insights into further developments in physical RC.

Some physical reservoirs are promising for developing next-generation machine learning hardware devices and chips. Such hardware-based RC can significantly speed up data processing compared to software-based RC, and it is often motivated by the need to reduce the power consumed by machine learning hardware. In particular, the RC framework using a simple readout is suited to low-cost real-time computation, which is not achievable with other machine learning frameworks based on iterative learning algorithms leading to high learning cost. Currently, many online services rely on cloud computing \citep{armbrust2010view} where tasks are performed at computers far from devices of end users on the Internet. Owing to the ever-growing amount of data at network edges, increasing communication latency is becoming a bottleneck for high-speed cloud computing. An emerging alternative computing paradigm for reducing the latency is edge computing \citep{shi2016edge} where tasks are performed at computers close to devices of end users and sensors, preferably in real time. Machine learning hardware for real-time temporal data processing, such as RC hardware, enables efficient edge computing, and therefore, is expected to become increasingly significant in the future. On the other hand, physical constraints make it difficult to optimize the conditions of reservoirs in physical RC. Many issues remain to be addressed in order to realize efficient physical RC systems and devices for practical applications. Our review introduces potential candidates of physical reservoirs for such devices.
 


The remainder of this review is organized as follows. In Sec.~\ref{sec:rc}, we briefly describe the basic concept of RC and its recent trends. In the subsequent sections, we introduce different types of physical RC systems, including RC based on well-known dynamical systems models (Sec.~\ref{sec:ds}), electronic RC (Sec.~\ref{sec:electronic}), photonic RC (Sec.~\ref{sec:photonic}), spintronic RC (Sec.~\ref{sec:spintronic}), mechanical RC (Sec.~\ref{sec:mechanical}), biological RC (Sec.~\ref{sec:biological}), and others (Sec.~\ref{sec:others}). Finally, we discuss the current issues and future research directions in Sec.~\ref{sec:discussion}.


\section{Reservoir computing (RC) \label{sec:rc}}

First, we outline the fundamental concept of RC in Sec.~\ref{subsec:framework}. For a more detailed background and overview of conventional RC, readers may refer to the following well-organized articles: the survey paper \citep{lukovsevivcius2009reservoir}, the review papers \citep{schrauwen2007overview,lukovsevivcius2012reservoir,scardapane2017randomness}, and the special issue on ESNs and LSMs in {\it Neural Networks} \citep{jaeger2007special}.
Then, we discuss recent trends in RC in Sec.~\ref{subsec:trend}.

\subsection{Basic framework \label{subsec:framework}}


Since the 1980s, RNNs have been used for temporal/sequential pattern recognition. They are characterized by feedback (recurrent) connections for generating history-dependent dynamical responses to external inputs. These models are described as non-autonomous dynamical systems. Although a special type of RNNs without external inputs (the so-called Hopfield networks \citep{hopfield1982neural}) have also been widely studied \citep{tanaka2019spatially}, we will not treat such models represented as autonomous dynamical systems in this review. The two major gradient-based learning algorithms forgeneral RNNs with external inputs are backpropagation through time (BPTT) \citep{rumelhart1985learning,werbos1990backpropagation} and real-time recurrent learning (RTRL) \citep{williams1989learning,doya1998recurrent}. These classical methods for training RNNs are discussed in a tutorial on RC \citep{jaeger2002tutorial}. In BPTT, an RNN is unfolded in time and regarded as an FNN with shared weights. Then the FNN is trained with the standard backpropagation algorithm \citep{rumelhart1985learning}. In practice, a variant of this algorithm, truncated BPTT, is performed using only a finite history of data to adaptively update the trained model and save computational cost. Therefore, this method has a difficulty in learning long-term dependencies of sequential data. On the other hand, RTRL shows excellent performance in online learning, but its time complexity (i.e. computational time) is high. Most RNN-based methods, including long short-term memory (LSTM) networks \citep{hochreiter1997long}, use one of the above-mentioned algorithms or a combination of them. 


In the early 2000s, ESNs \citep{jaeger2001echo,jaeger2004harnessing} and LSMs \citep{maass2002real,maass2011liquid} were independently proposed as seminal RC models. They are different from conventional RNNs in that the weights on the recurrent connections in the reservoir are not trained but only the weights in the readout are trained \citep{schrauwen2007overview,lukovsevivcius2009reservoir}. The nonlinear mapping of an input signal into a high-dimensional space in the reservoir is effective for pattern analysis of the input information as in the kernel method \citep{hofmann2008kernel}. To apply a simple machine learning method to the readout, the reservoir should be appropriately designed in advance. The characteristics of the two above-mentioned models, ESNs and LSMs, are briefly summarized below.

The ESN model was proposed by Jaeger \citep{jaeger2001echo,jaeger2004harnessing,Jaeger:2007}. This model uses an RNN-based reservoir consisting of discrete-time artificial neurons (Fig.~\ref{fig:rc}(a)). When the feedback from the output to the reservoir is absent, the time evolution of the neuronal states in the reservoir is described as follows \citep{jaeger2001echo}:
\begin{eqnarray}
\mathbf{x}(n) &=& \mathbf{f}(W^{\rm in} \mathbf{u}(n) + W \mathbf{x}(n-1)), \label{eq:esn_x}
\end{eqnarray}
where $n$ denotes discrete time, $\mathbf{x}(n)$ is the state vector of the reservoir units, $\mathbf{u}(n)$ is the input vector, $W^{\rm in}$ is the weight matrix for the input-reservoir connections, and $W$ is the weight matrix for the recurrent connections in the reservoir. The function $\mathbf{f}$ represents an element-wise activation function of the reservoir units, which is typically a sigmoid-type activation function. Equation~(\ref{eq:esn_x}) represents a non-autonomous dynamical system forced by the external input $\mathbf{u}(n)$. The output is often given by a linear combination of the neuronal states as follows:
\begin{eqnarray}
\mathbf{y}(n) &=& W^{\rm out} \mathbf{x}(n), \label{eq:esn_y}
\end{eqnarray}
where $\mathbf{y}(n)$ is the output vector and $W^{\rm out}$ is the weight matrix in the readout. In supervised learning, this weight matrix is trained to minimize the difference between the network output and the desired output for a certain time period (see Sec.~8 of \citep{lukovsevivcius2009reservoir} for details). The performance of the ESN depends on the design of the RNN-based reservoir. In order to approximate a teacher output signal, the RNN-based reservoir must have the {\it echo state property}, whereby it asymptotically eliminates any information from the initial conditions \citep{jaeger2001echo}. It is empirically observed that the echo state property is obtained for any input if the spectral radius (i.e. the maximum eigenvalue of $W$) is adjusted to be smaller than unity. A practical guide for reservoir design in ESNs can be found in \citep{lukovsevivcius2012practical}. The echo state property is essentially the same as the fading memory property \citep{boyd1985fading,matthews1993approximating}. An input/output system (or a filter) for temporal signals has fading memory when the outputs corresponding to inputs that are close in the recent past are close even if those inputs are very different in the distant past. Recently, it was theoretically shown that ESNs have universal approximation ability in terms of discrete-time fading memory filters with uniformly bounded inputs defined on negative infinite times \citep{grigoryeva2018echo}. 
The LSM was proposed by Maass et al. \citep{maass2002real,maass2011liquid} to explore the computational capability of neural microcircuits in the brain \citep{maass2004computational}. The purpose of LSMs is to develop biologically relevant learning models using spiking neural networks (SNNs) with recurrent architectures. The architecture of the LSM is similar to that shown in Fig.~\ref{fig:rc}(a) and the reservoir units are typically given by excitatory and inhibitory spiking neurons. Although the units are principally modeled with leaky integrate-and-fire (LIF) neurons, other biologically plausible spiking neuron models can also be used \citep{wojcik2007liquid,grzyb2009model}. The topology and connectivity of the RNN in the LSM follow the constraints of biological neural networks. Specifically, the probability that two neurons are connected depends on the distance between their positions. Such a reservoir is often called a {\it liquid} and the LSM operation is called {\it liquid computing} because it is similar to excitable media exhibiting ripples in response to external stimulation inputs. The reservoir dynamics is generally described as follows \citep{maass2002real}:
\begin{eqnarray}
x^M(t) &=& (L^M u)(t), \label{eq:lsm_x}
\end{eqnarray}
where $t$ denotes continuous time, $x^M$ is the reservoir state (neuronal activation patterns), $u(\cdot)$ is the input encoded as a spike sequence, and $L^M$ is the filter for transforming the input into the reservoir state. The output is given by 
\begin{eqnarray}
y(t) &=& f^M(x^M(t)), \label{eq:lsm_y}
\end{eqnarray}
where $y(t)$ is the output and $f^M$ is a memory-less readout map. A simple machine learning algorithm or a biologically plausible learning rule can be adopted to train the readout map. The LSMs can incorporate new findings about biological mechanisms of information processing in the brain.
It was shown that any given time-invariant filter (a transformation from $u(\cdot)$ to $y(\cdot)$) with the fading memory property can be approximated by LSMs to any degree of precision, if $L^M$ is chosen from a class of time-invariant filters with fading memory that has a point-wise separation property and $f^M$ is chosen from a class of functions that satisfies an approximation property \citep{maass2002real,maass2004computational}. Recent studies have treated the universal approximation property of LSMs and other RC systems in a more mathematically rigorous way \citep{grigoryeva2018echo,gonon2018reservoir}.

\subsection{Recent trends \label{subsec:trend}}

The number of studies on RC has been rapidly increasing in recent years. In this subsection, we discuss the recent trends in RC studies from several viewpoints, including applications, methods, and physical realizations.

First, RC has been successfully applied to many practical problems involving real data. One of the reasons for this success is that researchers outside the neural network community have recognized the advantages of RC. The simplicity of the training method in RC is attractive for non-expert developers. Table~\ref{tab:application} lists examples of subjects that have been addressed using RC. Most of these studies are involved in machine learning applications, such as pattern classification, time series forecasting, pattern generation, adaptive filtering and control, and system approximation. In particular, RC meets the demands for low training cost and real-time processing in these applications. Some benchmark tasks related to these applications are listed in Table~\ref{tab:benchmark}. The input and output information for an RC system are determined depending on the task. In pattern classification tasks, the input is a time series and the output is a discrete value (label) representing a pattern class. More specifically, in a spoken digit recognition task, the input is a sound signal corresponding to one of ten different utterances of the digits from zero to nine and the output is one of the ten digits. The goal of this task is to output the correct digit number from a sound signal of an unknown digit. To remove unnecessary information and noise, the original sound signal is typically transformed into a feature value such as mel-frequency cepstrum coefficient (MFCC) in a preprocessing step and then given to the reservoir. Although RC is suited for temporal pattern recognition, it can be applied to image recognition by transforming an image into a sequence of pixel values. As RC can deal with any sequential data in principle, further expansion of its application fields is widely expected.

\begin{table}[t]
\begin{center}
\caption{Examples of subjects in RC applications.}
\begin{tabular}{|l||l|}
\hline
Category & Examples  \\
\hline
\hline
Biomedical & EEG, fMRI, ECG, EMG, heart rates, biomarkers, \\
& BMI, eye movement, mammogram, lung images. \\
\hline
Visual & Images, videos. \\
\hline
Audio & Speech, sounds, music, bird calls. \\
\hline
Machinery & Vehicles, robots, sensors, motors, compressors, \\
 & controllers, actuators. \\
\hline
Engineering & Power plants, power lines, renewable energy, \\
 & engines, fuel cells, batteries, gas flows, diesel oil, \\
 & coal mines, hydraulic excavators, steam generators, \\
 & roller mills, footbridges, air conditioners. \\
\hline
Communication & Radio waves, telephone calls, Internet traffic. \\
\hline
Environmental & Wind power and speed, ozone concentration, \\
 & PM2.5, wastewater, rainfall, seismicity. \\
\hline
Security & Cryptography. \\
\hline
Financial & Stock price, stock index, exchange rate. \\ 
\hline
Social & Language, grammar, syntax, smart phone. \\
\hline
\end{tabular}
\label{tab:application}
\end{center}
\end{table}

Second, many variants of RC models have been proposed to improve the performance of the original ones. New RC models have been devised by using new architectures such as multiple reservoirs \citep{gallicchio2017deep,malik2017multilayered,akiyama2019analysis} and evolving reservoirs \citep{qiao2017growing}, combining RC with other feature extraction methods such as untrained convolutional neural networks \citep{tong2018reservoir} and reinforcement learning \citep{murakami2015seeing,chang2018distributive}, incorporating new learning algorithms such as the FORCE learning \citep{sussillo2009generating} and its variants \citep{pyle2018model}, and/or employing a diversity of reservoir elements \citep{xia2015quaternion,tanaka2016exploiting,inubushi2017reservoir}. In addition, theoretical studies have provided a deeper understanding of the relationship between the computational performance of RC and the dynamics of reservoirs, in terms of nonlinear dynamical systems theory, information theory, and statistical theory. A comprehensive overview of these studies is beyond the scope of this review.  



\begin{table}[t]
\begin{center}
\caption{Applications and related benchmark tasks of RC.}
\begin{tabular}{|l|l|}
\hline
Applications & Benchmark tasks \\
\hline
\hline
Pattern classification & Spoken digit recognition \citep{verstraeten2005isolated} \\
 & Waveform classification \citep{paquot_optoelectronic_2012}\\
 & Human action recognition \citep{soh2012iterative} \\
 & Handwritten digit image recognition \citep{jalalvand2015real} \\
\hline
Time series forecasting & Chaotic time series prediction \citep{jaeger2001echo} \\
 & NARMA time series prediction \citep{jaeger2003adaptive} \\ 
\hline
Pattern generation & Sine-wave generation \citep{jaeger2002tutorial} \\
 & Limit cycle generation \citep{hauser2012role} \\
\hline
Adaptive filtering and control & Channel equalization \citep{jaeger2004harnessing}\\
\hline
System approximation & Temporal XOR task \citep{bertschinger2004real} \\
 & Temporal parity task \citep{bertschinger2004real}\\
\hline
Short-term memory & Memory capacity \citep{jaeger2001short} \\
\hline
\end{tabular}
\label{tab:benchmark}
\end{center}
\end{table}

Finally, physical realizations of RC models have attracted considerable attention. A straightforward method is to implement RNNs using neural network hardware or neuromorphic computing techniques. Another method is to employ other dynamical systems instead of RNNs. Any dynamical system has the potential to serve as a reservoir if it can exhibit dynamical responses to inputs. Such reservoirs were previously regarded as ``exotic'' ones \citep{lukovsevivcius2009reservoir}, but the number of studies on physical RC has been rapidly increasing. Various physical reservoirs have been proposed using different types of physical systems, substrates, and devices. Some physical RC systems are aimed at developing energy-efficient machine learning hardware and others are at exploring natural computing based on novel substrates.

There are several requirements for a physical reservoir to efficiently solve computational tasks. (i) {\it High dimensionality} is necessary to map inputs into a high-dimensional space. This property facilitates the separation of originally inseparable inputs in classification tasks and allows reading out spatiotemporal dependencies of inputs in prediction tasks. The dimensionality is related to the number of independent signals obtained from the reservoir. 
(ii) {\it Nonlinearity} is necessary for a reservoir to operate as a nonlinear mapping. This property allows inputs that are not linearly separable to be transformed into those that are linearly separable in classification tasks. It is also useful for effectively extracting nonlinear dependencies of inputs in prediction tasks. (iii) {\it Fading memory} (or short-term memory) \citep{boyd1985fading,maass2002real,maass2004fading} is necessary to ensure that the reservoir state is dependent on recent-past inputs but independent of distant-past inputs. It is also referred to as the echo state property, indicating that the influence of past inputs on the current reservoir states and outputs asymptotically fades out \citep{jaeger2001echo,yildiz2012re}. Such a property is particularly important for representing sequential data with short-term dependencies. (iv) {\it Separation property} is required to separate the responses of a reservoir to distinct signals into different classes. On the other hand, a reservoir should be insensitive to unessential small fluctuations, such as noise, so that similar inputs are classified into the same class. Therefore, when a system parameter variation causes a transition between non-chaotic and chaotic regimes, it is often recommended that the parameter be set close to the transition point (the so-called edge of chaos \citep{bertschinger2004real,legenstein2007edge}) where the transformation by a reservoir is neither very expanding nor very contracting.

The responses of physical RC systems are used to train a readout that is realized using physical devices or software-based computations. Linear regression or another simple machine learning algorithm is used in the readout of ESN-type RC \citep{lukovsevivcius2009reservoir}, while a perceptron-like local learning rule or a synaptic plasticity-based one is used for the readout neurons in LSM-type RC \citep{maass2002real}.



\begin{figure}[t]
\begin{center}
\includegraphics[width=0.7\hsize]{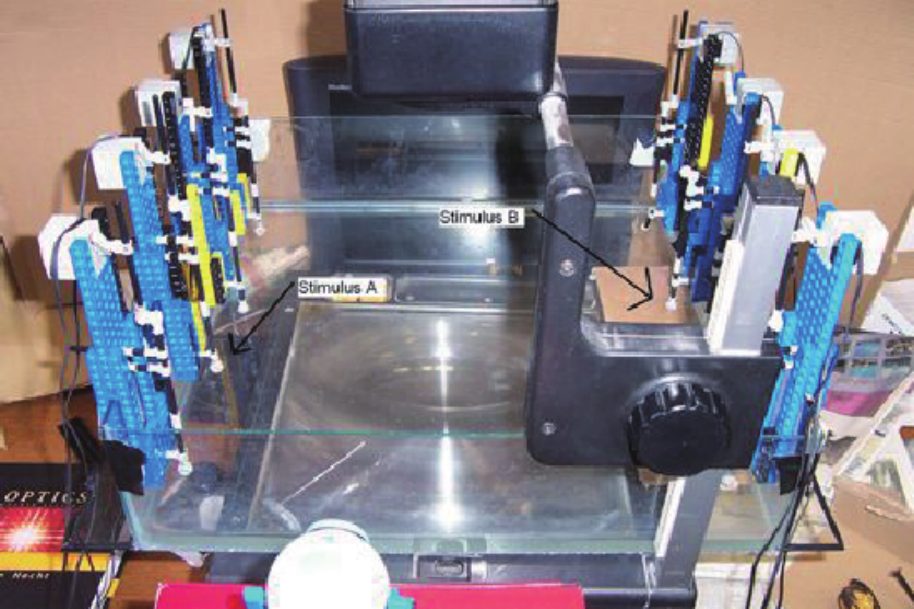}
\caption{A fluidic RC where the reservoir is the water in a bucket \citep{fernando2003pattern}. Figure reproduced with permission from Springer Nature.}
\label{fig:water}
\end{center}
\end{figure}

An intriguing example of physical RC in an early study is a fluidic RC with water in a bucket for pattern recognition as shown in Fig.~\ref{fig:water} \citep{fernando2003pattern}. The input signals are transmitted to electric motors that generate ripples on the water surface, and the ripples are recorded using a video camera. From the recorded signals, a software-based readout is trained. The performance of the liquid computer is demonstrated in an XOR task and a spoken digit recognition task. Subsequently, many dynamical systems models and physical systems have been employed as potential reservoirs. We classify these RC systems, substrates, and devices, depending on the type of the physical phenomenon in the reservoir and review the individual reservoirs in the following sections.

\section{Dynamical systems models for RC \label{sec:ds}}

In this section, we review several types of reservoirs based on well-known nonlinear dynamical systems models, including delayed dynamical systems (Sec.~\ref{subsec:delay}), cellular automata (Sec.~\ref{subsec:ca}), and coupled oscillators (Sec.~\ref{subsec:co}).

\subsection{Delayed dynamical systems \label{subsec:delay}}
\begin{figure}[t]
\begin{center}
\includegraphics[width=0.8\hsize]{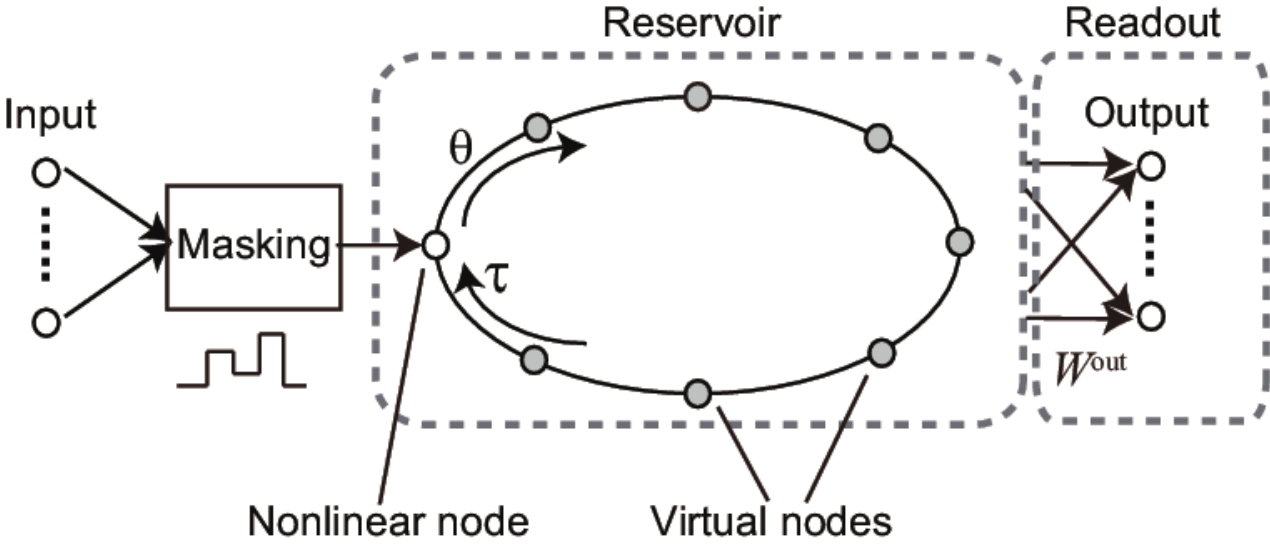}
\caption{RC using a single nonlinear node reservoir with time-delayed feedback \citep{appeltant2011information}.}
\label{fig:single-node}
\end{center}
\end{figure}

The reservoirs in ESNs and LSMs generate high-dimensional signals using a network of interacting neuron nodes, which are regarded as a special class of high-dimensional dynamical systems. Another way to generate high-dimensional patterns is to use a time-delayed dynamical system, as described in the following form \citep{lepri1994high}:
\begin{equation}
\frac{{\rm d} x(t)}{{\rm d}t} = F(t,x(t), x(t-\tau)), \label{eq:delay_diff}
\end{equation}
where $t$ represents continuous time, $x$ is the state variable, $F$ is a function determining the flow of this system, and $\tau>0$ is the delay period. This system is capable of exhibiting rich nonlinear behavior including periodic oscillations and deterministic chaos depending on the system parameter setting. In the first proposal of a single-node reservoir with delayed feedback \citep{appeltant2011information}, the reservoir was implemented using electronic circuits with a feedback loop (see Sec.~\ref{subsec:analog} for details). As shown in Fig.~\ref{fig:single-node}, the input signal is time-multiplexed by a mask function \citep{appeltant2014constructing} and fed to the single nonlinear node. The virtual nodes are set at $N$ time points that equally divide the delay period $\tau$. The time interval between two consecutive nodes is $\theta \equiv \tau/N$. The states at these virtual nodes, $x(t-(N-i)\theta)$ for $i=1,\ldots,N$, are used as the reservoir state at time $t$ and then fed to the output layer through weighted connections. These connection weights are trained in the readout. The system was successfully applied to the spoken digit recognition task and the nonlinear autoregressive moving average (NARMA)-10 time series prediction task. 

The architecture of the single-node reservoir with delayed feedback was extended in two ways \citep{ortin2017reservoir}. One is an ensemble of two separate time-delayed reservoirs whose outputs are combined at the readout. The other is a circular concatenation of the delay lines of two reservoirs, forming a longer delay line. These extended architectures were shown to achieve better performance, faster processing speed, and higher robustness than the single-node reservoir. An extensive amount of work has been performed on single-node reservoirs with delayed feedback \citep{brunner2018tutorial}.

The simplicity of the single-node reservoir with delayed feedback is advantageous for physical implementation compared with network-based reservoirs consisting of a large number of nodes. In fact, single-node reservoirs have been widely employed for electronic RC (Sec.~\ref{sec:electronic}) and photonic RC (Sec.~\ref{subsec:timedelay}).

\subsection{Cellular automata \label{subsec:ca}}
A cellular automaton (CA) is a simple dynamical systems model where both state and time are discrete \citep{wolfram2018cellular}. The discrete states on the cells are updated according to a given (local) evolution rule. Depending on the rule, the CA can exhibit rich behavior, including ordered, critical (or the edge of chaos), and chaotic dynamics, in spite of its simplicity. It is heuristically conjectured that the computational capability of CA is maximized at the edge of chaos. 

\begin{figure}[t]
\begin{center}
\includegraphics[width=0.7\hsize]{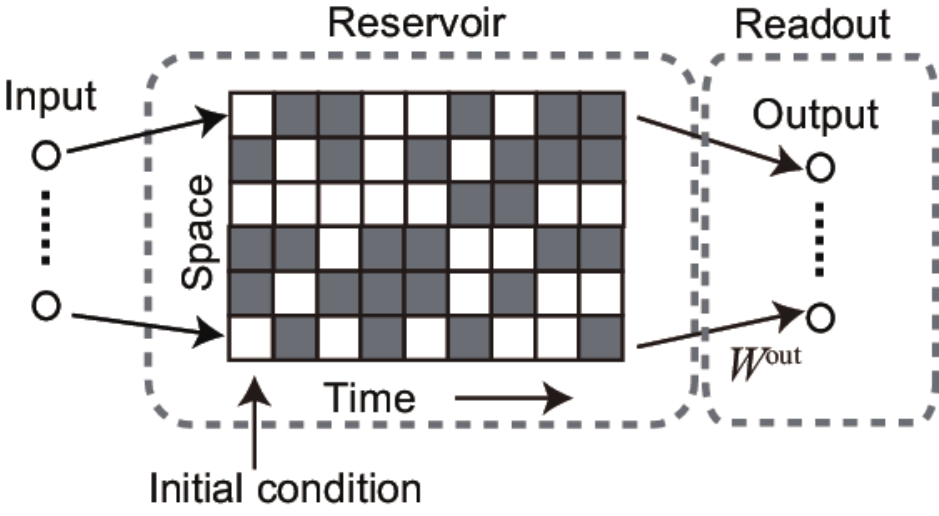}
\caption{RC using cellular automata \citep{yilmaz2015symbolic}.}
\label{fig:ca}
\end{center}
\end{figure}

This conjecture has been confirmed in a numerical study on RC based on a random Boolean network, which is an extended version of CA \citep{snyder2013computational}. Other studies employed CA-based reservoirs as shown in Fig.~\ref{fig:ca} \citep{yilmaz2014reservoir,yilmaz2015machine,yilmaz2015symbolic}. Binary or non-binary inputs are randomly mapped onto the initial states of CA through an encoding procedure. According to the predefined evolution rule, CA exhibits nonlinear dynamical behavior, through which the input data are projected onto an expressive and discriminative space. The entire state of the CA evolution is vectorized and used as a feature vector for processing in the readout. The CA reservoirs are binary in nature and suitable for symbolic computation including Boolean logic. The CA-based RC system can perform 5-bit and 20-bit temporal memory tasks, which require long short-term memory capability, with less computation compared to ESNs \citep{yilmaz2015machine}. The binary operations and simple update rules of CA reservoirs are advantageous for implementation with parallel hardware, such as field-programmable gate arrays (FPGAs) and graphics processing units (GPUs). In a recent study \citep{moran2018reservoir}, different CA evolution rules were tested for the MNIST handwritten character recognition by numerical simulations and the best rule giving the highest accuracy was specified as rule 90 \citep{martin1984algebraic}. Then, a CA-based reservoir with this rule was implemented with FPGA and applied to the MNIST task. Compared to other FPGA-based neural networks, the proposed method achieved competitive results in terms of accuracy, speed, and power dissipation. 

The architecture of the CA reservoir can be extended in several ways. For instance, a parallel loosely coupled architecture \citep{nichele2017reservoir} and a layered deep architecture \citep{nichele2017deep} have been proposed. The rules of CA can also be extended by employing two CA rules of different classes in the reservoir layer to satisfy two competing requirements of the CA reservoir, i.e., sensitivity to the current input and asymptotic insensitivity to past inputs \citep{mcdonald2017reservoir}. The first half of the CA reservoir is driven by a rule generating chaos or the edge of chaos for hyperdimensional projection of the inputs, while the second half is driven by a rule generating an ordered state for short-term memory.

\subsection{Coupled oscillators \label{subsec:co}}
Coupled nonlinear oscillators are ubiquitous in mechanical, chemical, electronic, biological, optical, spintronic, and quantum mechanical systems. They can be used as physical reservoirs. When each oscillator is described with a first-order ordinary differential equation (ODE), a system of $N$ coupled oscillators can be described in the following general form:
\begin{eqnarray}
\frac{{\rm d}x_i(t)}{{\rm d}t} &=& F(x_i(t))+G(x_1(t),\ldots,x_N(t)), \quad \mbox{for}~i=1,\ldots,N, \label{eq:coupled}
\end{eqnarray}
where $t$ represents continuous time, $x_i(t)$ is the state of oscillator $i$ at time $t$, $F$ is a function determining the dynamics of isolated oscillators, and $G$ is a coupling function. The coupled oscillator model consists of the term representing the dynamics of individual oscillators and the coupling term representing the interactions between oscillators.

\begin{figure}[t]
\begin{center}
\includegraphics[width=0.8\hsize]{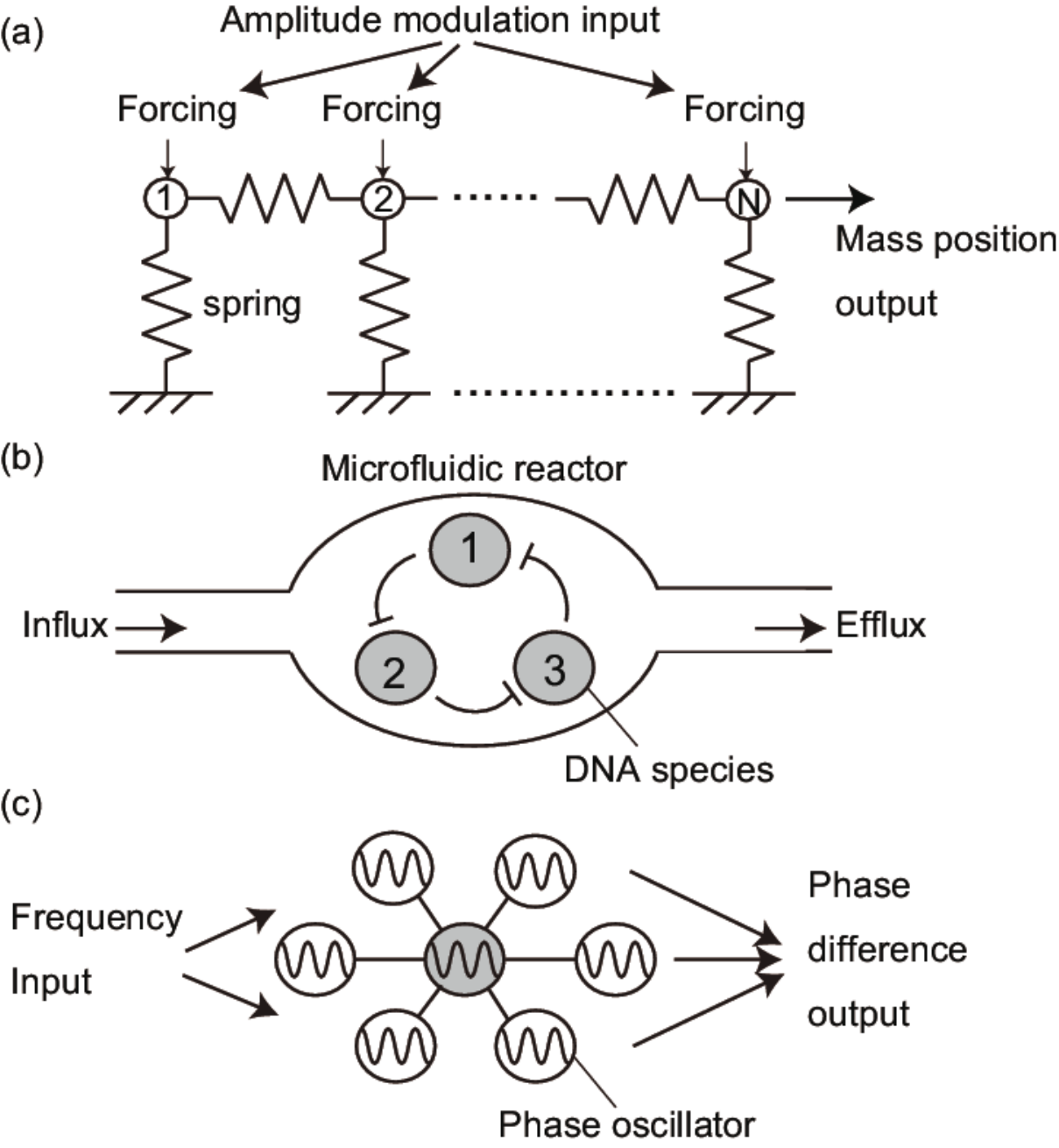}
\caption{RC using coupled oscillators. (a) A mechanical reservoir with coupled nonlinear oscillators \citep{coulombe2017computing}. (b) A DNA reservoir with coupled chemical oscillators \citep{goudarzi2013dna}. (c) A synchronization-based RC with coupled phase oscillators \citep{yamane2015wave}.}
\label{fig:co}
\end{center}
\end{figure}
 
A reservoir based on nonlinear mechanical oscillators has been proposed \citep{coulombe2017computing}. Such a reservoir is composed of multiple inertial masses arranged in a chain, which are coupled with their nearest neighbors by springs, as illustrated in Fig.~\ref{fig:co}(a). The masses are grounded by linear or nonlinear springs with damping. The entire equation of motion is described by coupled Duffing oscillators (second-order ODEs) driven by periodic forcing. The input signal is applied to the reservoir through amplitude modulation of the periodic forcing. The outputs from the reservoir are envelope signals containing only the low-frequency amplitude variations in the high-frequency mass position signals. The output signal is calculated by a linear combination of the weighted envelope signals. A reservoir with nonlinear mechanical oscillators can be compactly fabricated using microelectromechanical systems technology in an energy-efficient manner. The performance of such a reservoir computer was evaluated in a parity task and a spoken digit recognition task. Other mechanical reservoirs are reviewed in Sec.~\ref{sec:mechanical}.


Chemical reactions are often modeled with coupled chemical oscillators to reproduce their oscillatory behavior far from a steady state. A deoxyribonucleic acid (DNA) reservoir was designed with coupled deoxyribozyme-based oscillators at molecular scale as shown in Fig.~\ref{fig:co}(b) \citep{goudarzi2013dna}. The reservoir consists of different DNA species interacting via biochemical reactions in a microfluidic reaction chamber, which can be mathematically represented as coupled ODEs with state variables representing concentrations of the product molecules of the DNA species. The input signal is encoded as fluctuations in the influx of species into the reactor, and the reservoir state is monitored by fluorescent probes. A software-based readout is trained to map the oscillator dynamics to a target output. A chemical oscillator-based RC with three DNA species was applied to a temporal signal-tracking task.

When coupled nonlinear oscillators show limit cycle motions, they can be reduced to coupled phase oscillators under the assumption of weak interactions \citep{nakao2016phase}. After the reduction, the amplitude of the motion is eliminated and the dynamics is restricted to the phase domain. Coupled phase oscillators can exhibit rich dynamical behavior, including phase transition, clustering, and phase synchronization, which are available for RC. A phase-based RC encodes the input signal as phases of the oscillators and adopt phase synchronization for computation as shown in Fig.~\ref{fig:co}(c) \citep{yamane2015wave}. In this method, the phase coupling function is appropriately designed to perform function approximations. This approach based on synchronization can contribute to development of phase-based information processing \citep{parihar2017computing} and wave-based neuromorphic computing \citep{katayama2016wave}.

\section{Electronic RC \label{sec:electronic}}

RC systems implemented with electronic circuits and devices have been actively studied for developing machine learning devices with low training cost.   
Any existing ANN and neuromorphic circuits are available in principle as electronic reservoirs, but simpler configurations have been explored to reduce energy consumption, speed up computation, and cope with imperfection and noise in hardware. In this section, we start with single-node reservoirs implemented with analog circuits (Sec.~\ref{subsec:analog}) and then introduce RC systems implemented with FPGAs which are common reconfigurable hardware devices consisting of large arrays of simple configurable logic blocks and configurable interconnection structures (Sec.~\ref{subsec:fpga}). Subsequently, we review very-large-scale integrated circuit (VLSI) designs for RC devices (Sec.~\ref{subsec:vlsi}). Finally, we focus on memristive RC based on memristive units (Sec.~\ref{subsec:memristive}).

\subsection{Analog circuits \label{subsec:analog}}
\begin{figure}[t]
\begin{center}
\includegraphics[width=1.0\hsize]{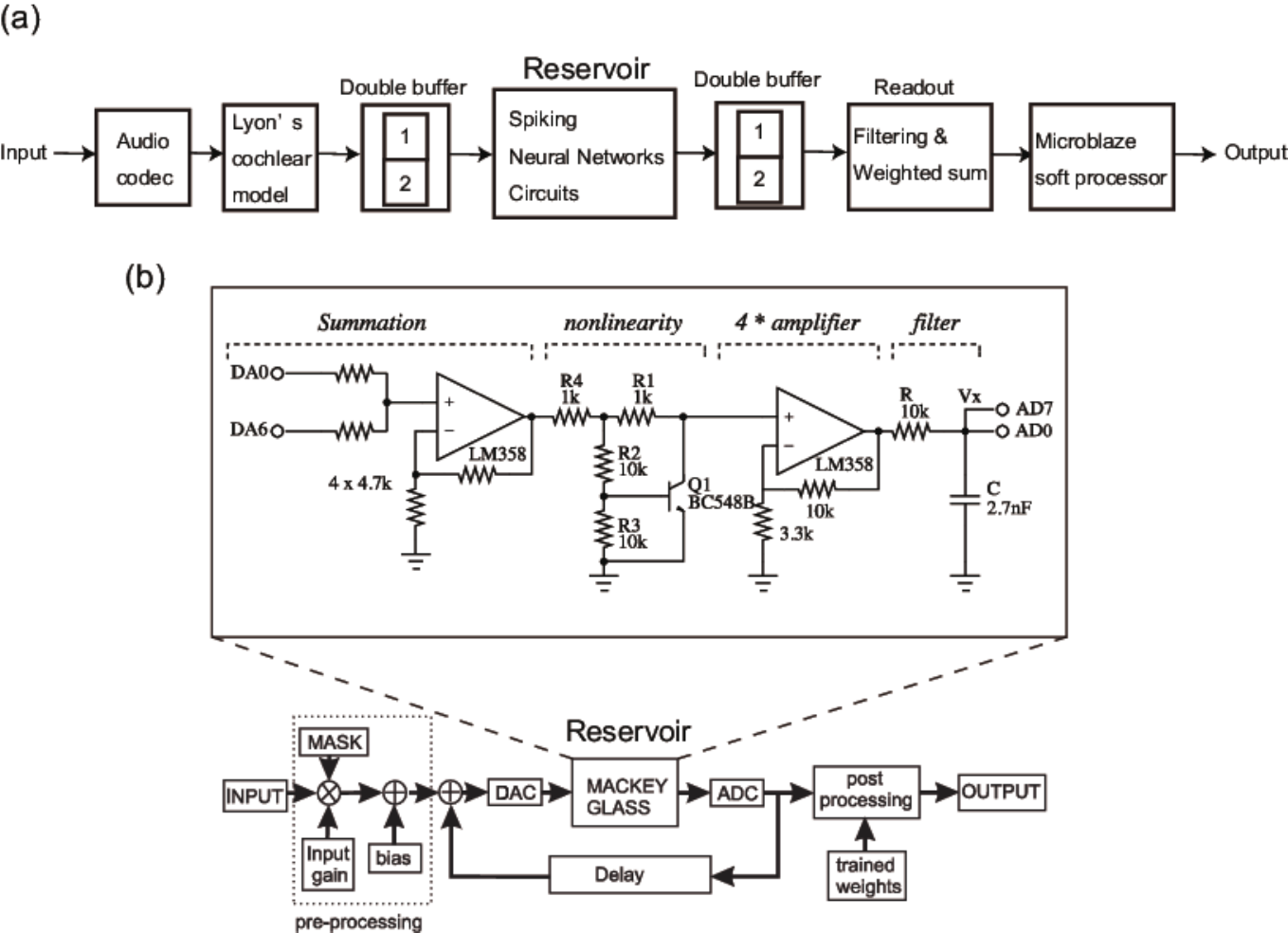}
\caption{(a) Electronically implemented RC system using a single-node reservoir based on the Mackey-Glass equation with delay \citep{soriano2015delay}. Figure reproduced and modified with permission from Springer Nature. (b) Architecture of an FPGA-based LSM system for speech recognition \citep{schrauwen2008compact}. Figure reproduced with permission from Elsevier.}
\label{fig:circuit}
\end{center}
\end{figure}

As described in Sec.~\ref{subsec:delay}, a single nonlinear node with delayed feedback works well as a reservoir where the input information is transformed into the states of the virtual nodes. The single-node reservoir imposes less hardware requirements compared to a network-type reservoir consisting of a large number of units and interconnections \citep{soriano2015minimal}. A nonlinear analog electronic circuit was implemented for constructing a single-node reservoir with a delay line, in combination with other digital hardware components for pre- and post-processing \citep{appeltant2011information,soriano2015delay}. The system architecture is schematically shown in Fig.~\ref{fig:circuit}(a). The input signal is time-multiplexed with a mask that defines the connection weights from the input to the virtual nodes on the delay line \citep{appeltant2014constructing} and kept to be positive by adding a bias voltage by digital processing. Then, the converted signal is injected into the single nonlinear node implemented with an analog Mackey-Glass nonlinear element circuit with a delayed feedback (the upper panel in Fig.~\ref{fig:circuit}(a)), corresponding to the following equation:
\begin{eqnarray}
\frac{{\rm d}x(t)}{{\rm d}t} &=& -x(t)+\frac{\eta (x(t-\tau)+\gamma I(t))}{1+(x(t-\tau)+\gamma I(t))^p}, \label{eq:MG}
\end{eqnarray}
where $t$ is dimensionless time, $x$ is the dynamical variable, $\tau$ is the delay in the feedback loop, $\eta$ is the feedback strength, $\gamma$ is the input scaling, $I(t)$ is the external input current, $p$ is the parameter for tuning the nonlinearity. The states of the virtual nodes are linearly combined to produce the output in the digital postprocessing part. The training of the output weights are performed using a linear regression algorithm. The digital and analog parts are interfaced by digital-to-analog (DA) and analog-to-digital (AD) converters with 12-bits resolution. The proposed system was successfully applied to spoken digit recognition, memory capacity estimation, and time series prediction, by appropriately adjusting the feedback strength $\eta$. The effect of the quantization noise caused by AD and DA conversion on the computational performance was also investigated in comparison with numerical simulations. A recent study proposed an extended system composed of multiple delayed feedback reservoirs based on the Mackey-Glass circuit in a deep layer structure \citep{li2018deep}. The presented system was applied to time series prediction tasks with Santa Fe dataset and ECG signals. Another study considered a single-node reservoir implemented with the Chua's circuit driven by external input forcing, which exhibits chaotic dynamics, and applied it to non-temporal nonlinear tasks \citep{jensen2017reservoir}.  

On the other hand, a spike-based single-node reservoir with a delay loop was proposed with its analog implementation design \citep{zhao2016novel}. The information transmission is based on spike signals for power efficiency, instead of time-continuous analog signals requiring peripheral modules for signal conversion and amplification. The spike-based delayed feedback circuit demonstrated nonlinear transformation from input spike sequences to output spike sequences. A spike timing-dependent encoder for encoding analog inputs as temporal spike trains and a masking process were used to design an RC system with a spike-based single-node reservoir \citep{li2017analog}. Numerical experiments showed that the optimal mask for yielding high computational performance is different depending on the task.

\subsection{FPGAs \label{subsec:fpga}}

FPGAs have often been used to implement ANNs as their reconfigurability is suited for concurrent processing and adaptive weight updating in ANNs \citep{zhu2003fpga,omondi2006fpga}. In the context of RC, many FPGA implementations have been studied to realize reservoirs and/or readouts \citep{antonik2018application}. Network-based reservoirs can be constructed using a variety of components, such as binary neurons, sigmoid neurons, stochastic neurons, and spiking neurons. A significant issue is how to choose hardware-friendly network components and efficiently implement them.

Binary neurons are often employed to compose neural network circuits as they are suited to be handled in digital platforms. In an early study, a reservoir composed of stochastic bitstream neurons was implemented on an FPGA board to explore efficient hardware implementation \citep{verstraeten2005reservoir}. The stochastic bitstream neuron is based on stochastic computing where its state value is represented as a statistical characteristic of a sufficiently long bitstream, e.g. as a probability of ``1'' in a binary sequence \citep{bade1994fpga}. Whereas sigmoid neurons require a large number of multipliers and adders for updating their states, stochastic bitstream neurons enable to simplify hardware implementation of those arithmetic operations. The proposed system with a small-world network of 25 neurons was applied to a simple task of generating a phase-shifted version of the input sinusoidal signal. A similar reservoir based on stochastic logic was implemented on an FPGA board using a random network of neuron units \citep{alomar2014low}. The proposed system composed of 10 neurons was applied to a nonlinear time series prediction task. The above FPGA-based reservoir computers use offline batch learning. This limitation was overcome by the first FPGA implementation of a reservoir computer with an online learning algorithm \citep{antonik2015fpga}. The neuron units in this reservoir have a sinusoidal activation function and the linear readout provides a gradient descent algorithm for real-time learning. It was confirmed that the reservoir computer on the FPGA board has a significant advantage in the high-speed processing over the software-based reservoir implemented in a high-end laptop. It was demonstrated that the proposed system can successfully solve channel equalization problems with invariable channels and those with variable channels.

Implementations of recurrent SNNs for LSMs on FPGAs have been reported in many studies. The hardware LSMs can be regarded as a special form of neuromorphic hardware. The information is coded by temporal patterns in spike sequences, unlike the rate coding in analog neuron hardware. Efficient hardware architectures for implementing LSMs on an FPGA were explored for real-time information processing and applied to isolated digit speech recognition \citep{schrauwen2008compact}. The overview of the system architecture for speech processing is shown in Fig.~\ref{fig:circuit}(b). The speech signals are transformed into cochleograms by the Lyon cochlear ear model \citep{lyon1982computational} and then converted to spike signals via Bens Spiker Algorithm \citep{schrauwen2003bsa} before being fed to the SNN reservoir. The spikes generated in the SNN are filtered by a low-pass filter and sampled, and the weighted sum of the sampled signals is used to produce the system output in the postprocessing step. The SNN consists of 200 LIF neurons and the exponential synapse models \citep{gerstner2002spiking}. A novel implementation technique for digital SNNs was devised on the basis of serial arithmetic operations instead of parallel ones, in order to enhance scalability of the hardware device. Another study presented an efficient design and architecture for general-purpose FPGA-based LSMs based on LIF neuron models with a biologically plausible learning rule \citep{wang2015general}. As a method to enhance energy efficiency, a light mode where some neuron units in the SNN are powered off was considered for simple tasks. The proposed neuromorphic LSM was successfully applied to speech and image recognition tasks with four different datasets. Furthermore, it was shown that the stand-alone FPGA-based LSM with parallel processing can speed up the runtime and reduce the energy consumption using the firing-activity dependent power gating and approximate arithmetic, compared to the implementation on a general-purpose CPU \citep{wang2016liquid,wang2017energy}. Other FPGA implementations of LSMs were demonstrated using stochastic spiking neurons in a ring network \citep{alomar2016stochastic}, multiplier-less reconfigurable architectures \citep{ghani2015reconfigurable}, and a spike-time-dependent encoder \citep{yi2016fpga}. In a recent work, a novel digital neuromorphic architecture called a spiking temporal processing unit was proposed by incorporating temporal buffers in each spiking neuron to model arbitrary synaptic response functions \citep{smith2017novel}. This scheme was adopted to implement an LSM on an FPGA chip, and its pattern classification ability was evaluated in a spoken digit recognition task. In another FPGA-based study, improvements in cost and energy efficiency of hardware LSMs were achieved by developing a reservoir tuning method based on a hardware-friendly spike-timing-dependent plasticity (STDP) learning algorithm in speech and image recognition \citep{liu2018online}.

The remaining FPGA-based reservoirs are the single-node reservoirs with delayed feedback (see Secs.~\ref{subsec:delay} and \ref{subsec:analog}). An FPGA-based single-node reservoir with a delay loop was demonstrated using a Boolean logic element \citep{haynes2015reservoir}. The Boolean logic element was set to operate as an XOR gate and the delay loop was realized as a cascade of pairs of inverter gates. The states of the virtual nodes are collected and linearly combined to produce an output in an external computer. It was shown that, by appropriately adjusting parameters such as the length of the time delay, the transient dynamics of the single-node reservoir is useful for classification of short input patterns. In another study, the single nonlinear node was realized with a digitized version of the Mackey-Glass equation in Eq.~(\ref{eq:MG}) and the delay line was implemented with the random access memory block \citep{alomar2015digital}. The training of output weights is conducted on the FPGA board. The presented system was successfully applied to a waveform pattern classification task and a time series prediction task.

\subsection{VLSIs \label{subsec:vlsi}}
Some efforts have made to efficiently integrating reservoir circuits in VLSIs. In an early study, an electronic reservoir was implemented with a general-purpose ANN application-specific integrated circuit (ASIC) which is a mixed-mode hardware using analog computation and digital signaling \citep{schurmann2005edge}. This system consists of 256 McCulloch-Pitts binary neuron nodes and 33k analog synapses. The input and output information are given by binary sequences. Experimental results showed that the computational performance of the RC system in a temporal 3-bit parity task is maximized when the neural network exhibits critical dynamics at the edge of chaos, as in the case of previous results of software simulations \citep{bertschinger2004real}.

Neuromorphic approaches to RC based on spikes and pulses have recently attracted much attention. Prototypes of pulse-based information processors were fabricated on a printed circuit board and an ASIC, on which RC systems can be implemented \citep{petre2016neuromorphic}. These systems are based on asynchronous pulse processing that uses a time encoder for converting analog signals into pulse domain signals and an event-driven computing scheme for power efficiency. In another study, a digital design architecture for implementing LSMs on reconfigurable platforms was proposed for real-time processing of input data \citep{polepalli2016digital,polepalli2016reconfigurable}. The reservoir consists of LIF neurons with random connectivity having spatial locality. The readout layer consists of a two-layer perceptron and the output weights are adjusted based on a gradient descent method. The performance of the proposed model was numerically evaluated in epileptic seizure detection from EEG signals and user identification from walking patterns. Hardware-friendly readout implementation techniques have been focused on in some studies for energy-efficient learning in VLSI-based LSMs. Inspired by dendrites in biological neurons, a hardware-friendly readout architecture for LSMs and an efficient learning rule with adaptive network rewiring were proposed \citep{roy2014liquid}. The efficiency of the proposed architecture was demonstrated in a spike train classification task and an approximation task of retrieving the sum of firing rates of input spike trains. In another study, a biologically inspired local learning rule was presented for low-power VLSI implementation of LSMs to reduce hardware implementation costs \citep{zhang2015digital}. It was numerically shown that the overhead of hardware implementation can be reduced by the new learning rule in a speech recognition task. The tradeoff between hardware overhead and computational performance in hardware-implemented LSMs was discussed in a simulation study \citep{jin2017performance}.

\subsection{Memristive RC \label{subsec:memristive}}

This section focuses on reservoirs implemented with memristive circuits and devices. The main characteristic of a memristive element, differentiating from other fundamental circuit elements, is that its resistance changes with time depending on the current flow that has passed through it \citep{chua1971memristor,williams2014we}. The existence of a memristor (or memory resistor) was first predicted from theoretical consideration \citep{chua1976memristive} and after a long time physically realized using titanium dioxide (TiO$_2$) \citep{strukov2008missing}. Although there are some debates about whether the memristor is truly counted as a fundamental passive circuit element \citep{abraham2018case}, memristive devices indeed exist in reality and are promising for computing. We classify memristive reservoirs into two types: neuromemristive reservoirs consisting of both neuron circuits and memristor synapses (Sec.~\ref{subsubsec:neuromem}) and memristive reservoirs without neuron units (Sec.~\ref{subsubsec:mem}).

\subsubsection{Neuromemristive circuits \label{subsubsec:neuromem}}

Neuromorphic computing with non-von Neumann architecture has rapidly progressed in recent years. Neuromorphic devices/chips have been fabricated by different research groups \citep{walter2015neuromorphic}. Neuromemristive systems are a subclass of neuromorphic computing systems that use memristors to mimic the synaptic plasticity in biological neurons, where the memristor conductance corresponds to the synaptic weight \citep{indiveri2013integration,thomas2013memristor}. Nano-scale memristive synapses have considerable potential as elements of energy-efficient neuromorphic devices.


A memristor-based ESN with cellular neural network structures (i.e., locally connected arrays) was proposed using staircase memristor models \citep{yang2016investigations}. Numerical simulation showed that, due to the simple structure, the performance of the proposed model was successful but worse than that of the original ESN model with a random connection topology in a time series prediction task. Other ESN-type reservoirs were designed using memristor crossbar arrays \citep{merkel2014memristive,donahue2015design}. Memristor crossbar arrays are often used to implement direct synaptic connections in neuromorphic devices, because they are suitable for vector-matrix multiplication, and the conductances of memristive synapses can be adaptively updated by applying voltage pulses. Double crossbar arrays were used to realize recurrent connections in ESNs \citep{hassan2017hardware} and LSMs \citep{soures2017robustness}. The use of memristor crossbar arrays in the readout part of ESNs was proposed for a digital (or mixed signal) reservoir with a doubly twisted toroidal structure \citep{kudithipudi2016design}. A more general on-chip system using analog memristive nanosynapses was presented for emulating a reservoir computer and performing recognition tasks \citep{bennett2017spatio}.

\subsubsection{Memristive systems and devices \label{subsubsec:mem}}

Memristive systems and devices are capable of exhibiting nonlinear dynamics and responding to inputs in a history-dependent manner. Even without neuron units, memristors can exhibit nonlinear transformation of input signals. By exploiting these favorable properties for a dynamic reservoir, some studies have proposed reservoirs based on memristive systems and devices. There are various memristive devices having different current-voltage characteristics, but ideally, they can be formulated as memristive systems categorized into a special class of dynamical systems \citep{chua1971memristor,chua1976memristive}. A current-controlled memristor with time-invariant characteristic is generally described as follows:
\begin{eqnarray}
V &=& R(\mathbf{w},I)I, \label{eq:memristive_v}\\
\frac{{\rm d}\mathbf{w}}{{\rm d}t} &=& f(\mathbf{w},I), \label{eq:memristive_w}
\end{eqnarray}
where $t$ represents continuous time, $V$ is the voltage, $R$ is the time-varying resistance, $I$ is the current, and $\mathbf{w}$ is a vector representing the internal state of the system. The function $f$ determines how the internal state evolves depending on the input current.

\begin{figure}[t]
\begin{center}
\includegraphics[width=\hsize]{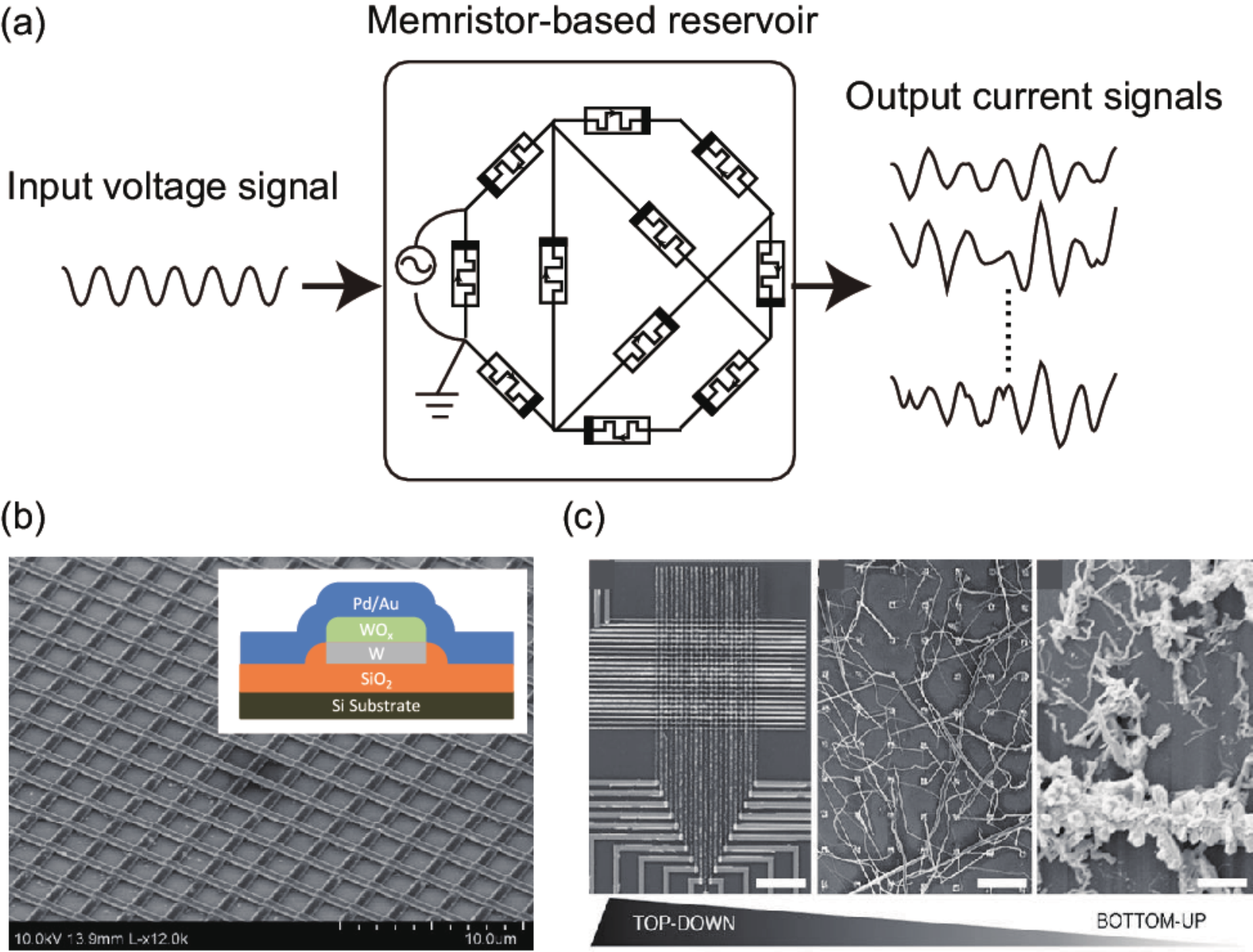}
\caption{(a) Schematic illustration of an RC system based on a memristor-based reservoir. (b) A fabricated memristor array where several cells are selected to work as a reservoir on the whole \citep{du2017reservoir}. The inset shows tungsten oxide (WO$_x$) memristor with a metal-insulator-metal structure. Figure reproduced from \citep{du2017reservoir}, licensed under CC-BY 4.0. (c) An atomic switch network device composed of self-assembled silver nanowires \citep{stieg2014self}. Figure reproduced with permission from Springer Nature.}
\label{fig:memristive}
\end{center}
\end{figure}

A network of memristors can be used as a reservoir to nonlinearly map an input signal into a high-dimensional feature space, as shown in Fig.~\ref{fig:memristive}(a). The first memristor-based reservoir was proposed in \citep{kulkarni2012memristor}. The proposed model was applied to a wave pattern classification task and an associative memory task, but the readout was a genetic algorithm which is not a linear algorithm. Therefore, the merit of the memristive reservoir was not clear from the results. As the connectivity of the memristors affects the computational performance of memristor-based RC, subsequent studies have extended the structure of the memristor network to a regular mesh structure for investigating the variation tolerance of RC \citep{burger2013variation} and to a hierarchical structure for enhancing the computational capability of RC \citep{burger2015hierarchical}. Moreover, the correlation between the computational capacity and the energy consumption of RC with random memristor networks was investigated to clarify its potential and limitations \citep{burger2015computational}. In the above-mentioned studies, the behavior of memristor-based reservoirs was computed using circuit simulators. On the other hand, a general mathematical model of memristor networks was explicitly formulated to promote theoretical and numerical analyses of memristive RC \citep{tanaka2017waveform}. This formulation is applicable to any memristor network if the memductance (memristor conductance) of a single memristor is expressed as a function of its magnetic flux as in the linear drift model \citep{strukov2008missing}. The numerical simulation showed that the variability in the memristive elements, which is generally thought to be undesired, can be beneficial if the network topology and the input scaling are appropriately selected. In \citep{carbajal2015memristor}, new memristor models with volatility were proposed to introduce the fading memory property into the standard non-volatile memristor model. Other memory devices, such as memcapacitors, are potential components of a network-type reservoir with low power consumption \citep{tran2017memcapacitive}. 

Several studies have demonstrated physical implementations of memristive reservoirs. A memristor-based reservoir device was recently fabricated as shown in Fig.~\ref{fig:memristive}(b) \citep{du2017reservoir}, where the reservoir is not a network of memristors but a group of independent memristors. An input signal is divided into multiple segments, and the segments are then separately transformed into output signals by the individual memristive devices. The collection of the output signals is used as the state of the entire reservoir. It was experimentally demonstrated that this RC device performs well in image recognition and time series prediction tasks. Memristive behavior is observed also in atomic switch networks (ASNs), which are aggregations of a number of silver nanowires interconnected with each other and formed via a thermodynamically driven self-organized growth process, as shown in Fig.~\ref{fig:memristive}(c) \citep{stieg2012emergent}. The RC device based on ASNs was applied to a waveform generation task \citep{sillin2013theoretical,stieg2014self}.

\section{Photonic RC \label{sec:photonic}}

In this section, the rich literature on photonic reservoir implementations is categorized and briefly summarized. An alternative discussion on this specific area can be found in \citep{van_advances_2017}. Reviews on topics related to photonic spike processing \citep{prucnal_recent_2016} and photonics for neuromorphic applications \citep{ferreira_progress_2017} have also been published recently. In this review, optical reservoirs are categorized as spatially distributed array reservoirs (Sec.~\ref{subsec:opticalnode}) and reservoirs with delay feedback (Sec.~\ref{subsec:timedelay}). On this last category, a detailed tutorial has recently been published \citep{brunner2018tutorial}.

\subsection{Optical node arrays \label{subsec:opticalnode}}

\begin{figure}[t]
\begin{center}
\includegraphics[width=0.9\hsize]{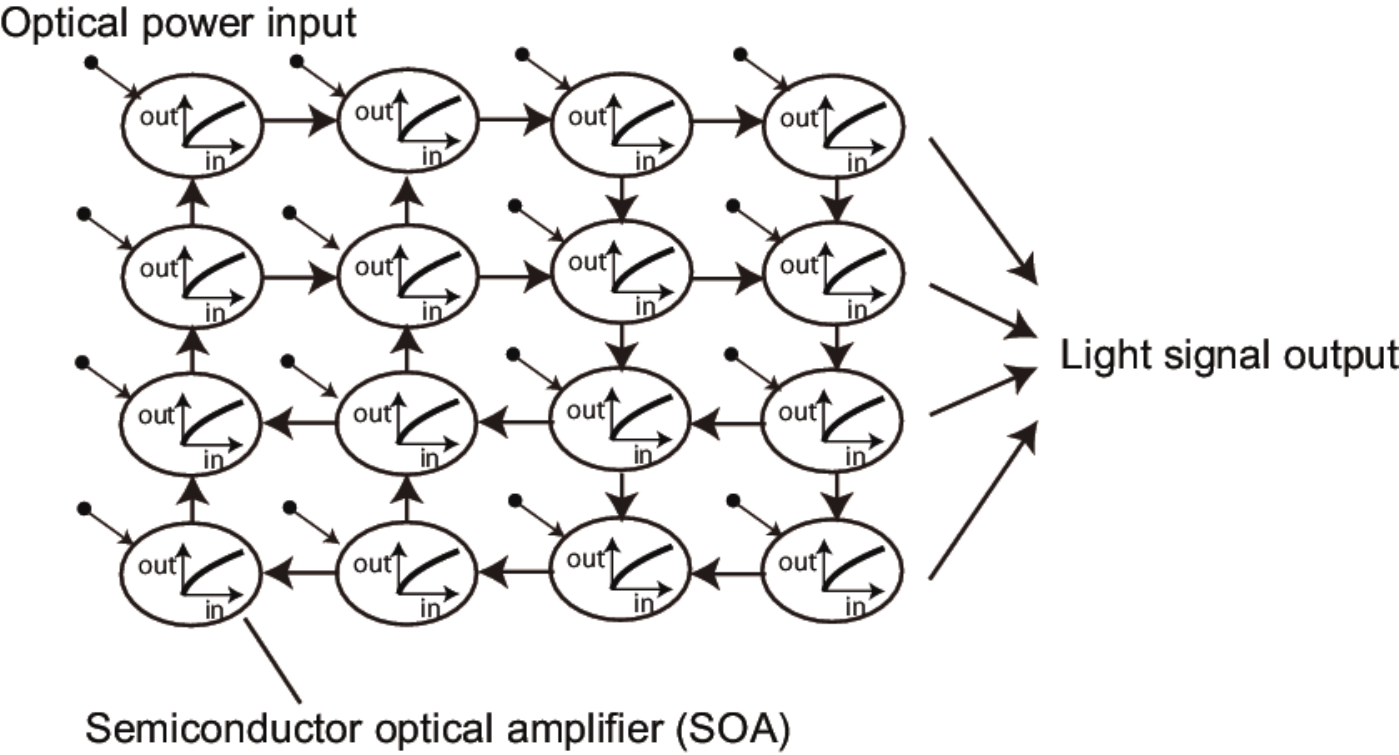}
\caption{A 4$\times$4 node array optical reservoir with a swirl connection pattern \citep{vandoorne_parallel_2011}.}
\label{fig:optical_nodearray}
\end{center}
\end{figure}

A photonic reservoir computer, with the potential advantages of low power consumption and extremely fast computation, was first proposed in 2008 and subsequently refined on the basis of numerical simulations \citep{vandoorne_toward_2008,vandoorne_parallel_2011}. The design is based on a chip-integrated device with single-mode waveguides. Semiconductor optical amplifiers (SOAs) are assembled in a 4$\times$4 array, with each node connected to a maximum of four neighbors in a swirl configuration, as shown in Fig.~\ref{fig:optical_nodearray}. The effect of non-uniformity of the nodes and the delay and phase shift between interconnections were studied. An experimental prototype was reported in 2014 \citep{vandoorne_experimental_2014}. Readout was performed at 11 of the 16 nodes by photodetectors providing nonlinearity, owing to the limitations imposed by optical loss. The signal input was inserted via a single node and the training of the reservoir and processing of the output were performed offline. Experimental demonstrations include a 2-bit XOR logical operator, header recognition, and spoken digit classification. 

A proposed related design is a platform in which the nodes are microring resonators providing nonlinear responses \citep{mesaritakis_micro_2013}. With a randomly interconnected 6$\times$6 array, successful operation for a pattern recognition task was numerically demonstrated. Another related study developed a node-based framework adapted to a coupled-mode theory to simulate a large number of cavities efficiently \citep{vaerenbergh_efficient_2015}. A drawback of photonic reservoirs is that a compact design results in short time delays; hence, the required input and readout operation rates may be too high for practical implementation. A digital mask modulation technique was reported to alleviate this problem and decrease the input signal rate by a factor of 40 \citep{schneider_using_2016}. There is a design choice for the number of nodes that are linked to the input signal. This aspect was recently studied, and it was found that the power efficiency was improved for an input signal fully connected to the array \citep{katumba_multiple-input_2017}. An intrinsic limitation of single-mode waveguide-integrated passive optical reservoirs is loss accumulation for a large number of nodes and long delay lines. A chip-integrated multimode photonic circuit with a low-loss optical Y-junction combiner was proposed, simulated, and applied to a header recognition task \citep{katumba_low-loss_2018}.

The concept of a device built instead on a photonic crystal platform was first proposed for a multiple superimposed oscillator waveform prediction task \citep{fiers_nanophotonic_2014}. Good simulation results were reported for a reservoir structure made of an array of resonator cavities for which there is bistability in the optical power. More recently, a reservoir built from photonic crystal cavities with multiple waveguide inputs and outputs was designed to exploit field mixing dynamics and was shown to exhibit memory up to 6 bits \citep{laporte_numerical_2018}. 

Another family of reservoirs with photonic nodes is based on free-space optics principles. A configuration with a diffraction grating and Fourier imaging was designed with randomly interconnected microring resonators and applied to an imaging pattern task \citep{mesaritakis_high-speed_2015}. Another reservoir with a diffractive optical element was recently described on the basis of an 8$\times$8 laser array and a spatial light modulator. Rich interaction dynamics was experimentally observed, and the potential for scaling up to a complex configuration with low power consumption was illustrated \citep{brunner_reconfigurable_2015,van_advances_2017}. Further modifications to this setup, including the use of a laser illumination field and digital micro-mirror device, led to the realization of an RNN with hundreds of nonlinear optical nodes applied to a time series prediction task \citep{bueno_reinforcement_2018}. A reservoir in which node interaction comes from the scattering of light passing through a microscopic slide with a simple layer of white paint was also proposed and applied to logical function operation, with signal input via a micro-mirror array \citep{dong2018scaling}.

\subsection{Time-delay systems \label{subsec:timedelay}}

\subsubsection{Opto-electronic and optical feedback with gain \label{subsubsec:oe}}

The most extensively studied implementation of an optical reservoir computer is a configuration that uses a single physical node with a time-delayed feedback signal \citep{appeltant2011information}, as explained in Sec.~\ref{subsec:delay}. The input signal is converted into a staircase waveform by a sample-and-hold procedure. A weight mask related to the number of virtual nodes in the feedback loop is applied to each symbol step. The weights of the output signal are calculated offline during the training procedure. In this section we describe this class of systems as well as related designs.

\begin{figure}[t]
\begin{center}
\includegraphics[width=0.7\hsize]{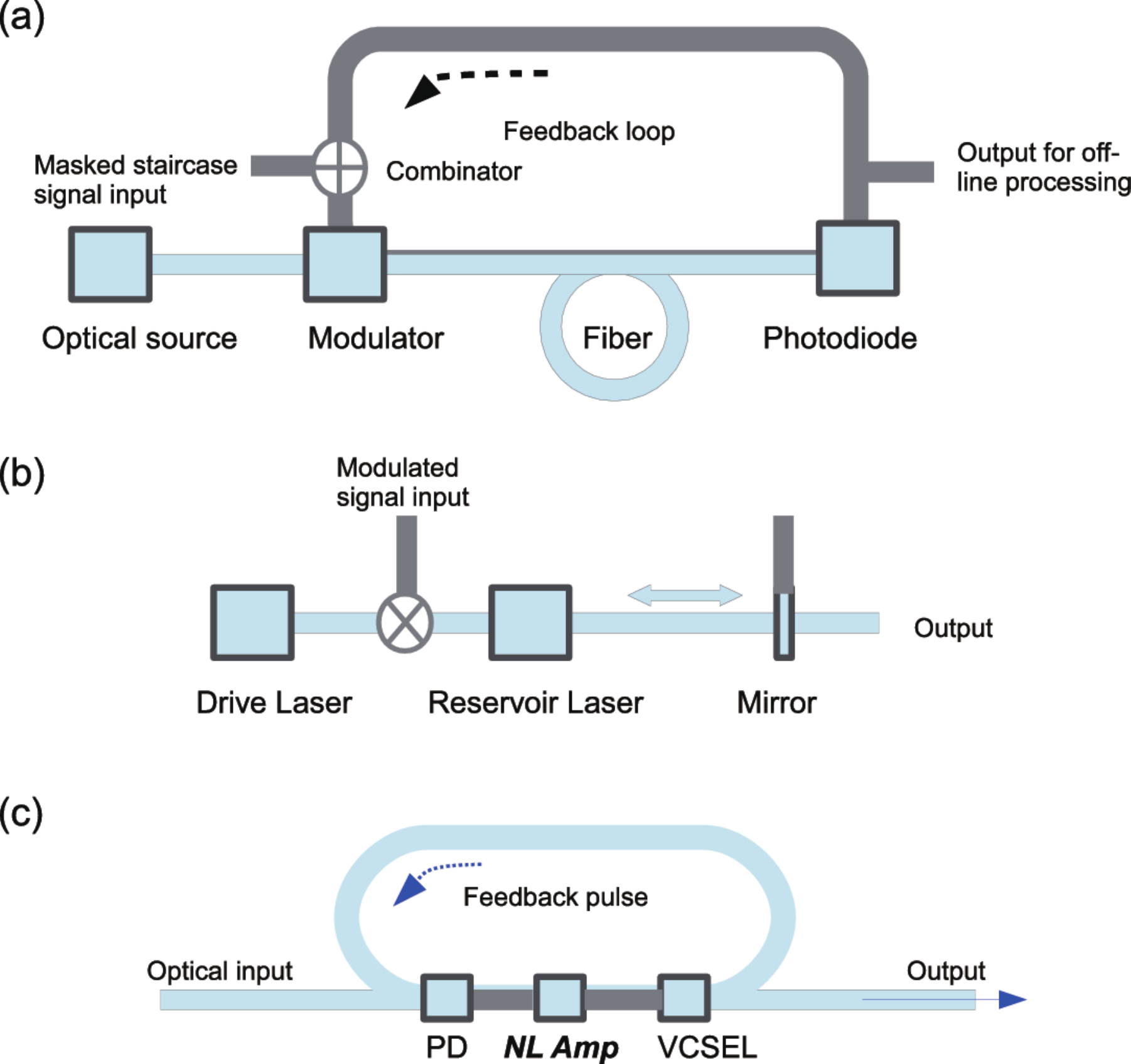}
\caption{Time-delay reservoir configuration examples. Blue and gray lines represent optical and electronic signals, respectively. (a) Opto-electronic feedback loop. (b) All-optical feedback loop with gain. (c) Passive optical feedback and opto-electronic node with gain. (d) Feedback into a laser cavity by a partially reflecting mirror.}
\label{fig:timedelay}
\end{center}
\end{figure}

In the first experimental demonstrations \citep{larger_photonic_2012,paquot_optoelectronic_2012}, the feedback loop was optoelectronic. Light from a continuously emitting laser source is sent to a modulator for information processing. The optical output reaching a photodiode is amplified and electronically combined with the input signal to drive the modulator, as shown in a generic form in Fig.~\ref{fig:timedelay}(a). Readout is performed either by sampling the feedback photodiode output or with an optical splitter collecting part of the light in the loop and a second photodiode. Demonstrations include generation of the NARMA equation of order 10, equalization of a nonlinear channel, and spoken digit recognition. In these pioneering demonstrations, the signal input rate was in the MHz range. 

A number of related published studies have reported different components, design optimization, and the addition of new functionality. A reservoir in which the input signal is fed by directly modulating the laser, node nonlinearity is achieved by a birefringent plate, and multiple feedback loops with different delay times for enhanced connectivity are realized with an FPGA, was reported \citep{martinenghi_photonic_2012}. Designs with an all-optical feedback in Fig.~\ref{fig:timedelay}(b) in which the electronics in the loop was replaced by a semiconductor optical amplifier and a fiber coupler \citep{duport_all-optical_2012} or a fiber amplifier, semiconductor saturable absorber, and fiber coupler \citep{dejonckheere_all-optical_2014} have also been demonstrated experimentally. 

A critical limiting factor for reservoir prototypes is the impact of noise on performance. This issue was investigated numerically and experimentally, and the importance of a preprocessing mask weight with multiple values, as opposed to binary values, was demonstrated \citep{soriano_optoelectronic_2013}. It was also shown that when an optoelectronic loop setup was used for a reservoir computer, the system could perform learning tasks even if the feedback was removed to obtain an extreme learning machine configuration if the response was slower than the masked input steps to provide some interaction between neighboring neurons \citep{ortin_unified_2015}. The system without feedback was found to be capable of performing a cancer type classification task with genetic microarray data. 

A chip-integrated device that combines physical optical nodes and time memory functions by connecting in series several ring resonators with a time-encoded input optical signal was also described \citep{zhang_integrated_2014} and is included here for completeness even though the feedback is not amplified. Performance for signal classification and chaotic time series prediction tasks was numerically investigated. In another scheme, a reservoir with a coherently-driven passive cavity was reported, in which the low optical loss led to good operation \citep{vinckier_high-performance_2015}. State-of-the-art results were obtained numerically and experimentally for a spoken digit recognition task. A parallel reservoir computer based on a fiber loop with frequency multiplexing of neurons was also reported \citep{akrout_parallel_2016}, where the realization of nonlinear channel equalization and isolated spoken digit recognition was simulated.

Another example of added functionality is the simultaneous computation of three independent tasks in a time-delay optical system with three so-called virtual reservoirs, as the system typically has an inherent bandwidth higher than the input signal rate \citep{duport_virtualization_2016}. To avoid electronic preprocessing of the input signal into a step waveform as an input mask, an implementation of a fully analogue reservoir with time delay was also proposed and only slight performance degradation compared to a step signal was observed \citep{duport_fully_2016}. At the output layer, the signal was split between a readout photodiode and a modulated portion applying the readout coefficients online. In another publication, the potential of an optoelectronic feedback reservoir computer for high speed processing was clearly demonstrated in the realization of a setup capable of performing a million-words-per-second classification task using fast electronics for input and output processing \citep{larger_high-speed_2017}. Phase modulators were used in the feedback loop, and the delay between the virtual nodes corresponded to an input signal rate of approximately 17 GHz.

An interesting application was shown in \citep{qin_numerical_2017}, namely the identification and classification of a packet header for switching in an optical network application. The low hardware requirements compared to a traditional neural network were emphasized. This concept was further extended to a system with two feedback loops to perform simultaneous recognition of packet headers for two optical channels \citep{zhao_simultaneous_2018}. An optoelectronic reservoir computer was also applied to the task of generating a long-range periodic time series and the emulation of a chaotic system \citep{antonik_brain-inspired_2017, antonik_random_2017,antonik2018application}. The underlying concept of this work was to feed the signal output processed by an FPGA back into the reservoir, in addition to the usual feedback signal. Periodic frequency and random patterns were successfully generated. 

Some recently published articles have reported the potential of novel schemes for online training. A new method based on gradient descent training with BPTT was recently proposed and experimentally shown with an optoelectronic feedback loop \citep{hermans_photonic_2015, hermans_optoelectronic_2015, hermans_embodiment_2016}. The scheme cannot be strictly labelled as a reservoir, because the input weights are optimized to enhance performance, but may have a significant impact on the development of future related systems. Results obtained for the TIMIT phoneme recognition task were similar to those reported for a fully offline trained RC with a larger number of nodes. By connecting an FPGA performing a gradient descent algorithm at the loop output, a device with a very low symbol error rate was also demonstrated for a distorted wireless signal recovery task, particularly for a high signal-to-noise ratio environment \citep{antonik_online_2017}. 

A multimode polymer waveguide-based design in which coupling and propagation losses are small has recently been proposed \citep{heroux2017polymer, Heroux_Optoelectronic_2018}. The feedback is all-optical and the nodes are opto-electronic, composed of photodiode, amplifier, and vertical cavity surface emitting laser (VCSEL) chip arrays, as shown in Fig.~\ref{fig:timedelay}(c). Simulations with this reservoir yielded promising results for signal recovery and nonlinear time series prediction tasks. As the gain is located in the neurons, there is no fundamental limitation on the aggregation of a large number of physical nodes in a future device. Moreover, as the optical path is on the centimeter scale with low loss and signals can cross each other, there is added flexibility, compared to a chip-integrated device, in the implementation of multiple relatively long delay feedbacks on a solid substrate in a compact format. 

\subsubsection{Optical feedback in a laser cavity \label{subsubsec:cavity}}

A second class of time-delayed optical reservoir computers has been investigated in recent years, in which the nonlinearity comes not from a discrete element in the loop but from the nonlinear response of a laser when a delayed signal re-enters its cavity. Light input and output with the same aperture is achieved by polarization rotation with a circulator component or another suitable mechanism. In the first reports \citep{brunner_parallel_2013, hicke_information_2013}, good results were experimentally obtained with a semiconductor laser for spoken digit recognition and chaotic time series prediction tasks. In another study that is related but not strictly categorized as a reservoir as there was no feedback component, a similar system based on a nonlinear resonant cavity was used to demonstrate optical vector-matrix product operations \citep{brunner_high-speed_2013}.

In an alternative configuration, a semiconductor ring laser was implemented as the nonlinear node, with the potential for on-chip integration and simultaneous processing of two tasks by modes with opposite propagation directions. Numerical studies have shown good results for chaotic time series prediction and nonlinear channel equalization tasks \citep{nguimdo_simultaneous_2015}. Delayed systems are sensitive to phase change, but in a subsequent study, it was explained that this problem can be alleviated with a modified readout layer \citep{nguimdo_reducing_2016}. 

In another related setup, information was sent into the laser cavity via phase modulation, and a partially transparent mirror split the laser output into a delayed component going back into the cavity and a transmitted component for readout \citep{nakayama_laser_2016}, as shown schematically in Fig.~\ref{fig:timedelay}(d). The efficiency of a chaotic input mask generated from a separate optical cavity subjected to feedback was compared to other digital and analog masking schemes and numerically studied for a chaotic time series prediction task. Performance improvement was observed, provided that the mask frequency was near the relaxation oscillation of the ring cavity laser. It was experimentally confirmed that a chaotic or colored-noise mask gives better results with a properly selected cut-off frequency \citep{kuriki_impact_2018}. In another study, numerical simulation of a reservoir in which the input signal was applied to a tunable Bragg reflector providing external feedback to the laser cavity was proposed, and the reservoir was applied to a waveform classification task \citep{takeda_photonic_2016}.

Another application was described in \citep{qin_optical_2016}, where a delayed optical feedback setup with a polarization circulator was adopted for optical packet header identification at 10 GB/s. The optimal feedback parameters were found and a low recognition error was obtained provided that the signal-to-noise ratio was above 15 dB. Calculations with a ring laser system for dual-channel packet header recognition have also been reported \citep{bao_recognition_2018}.

The optimal operation conditions for an all-optical feedback loop system were experimentally investigated in \citep{bueno_conditions_2017}. The effects of detuning of the frequency between an injection laser and the reservoir laser as well as the locking of the laser state were studied for a chaotic time series prediction task. Moreover, very recently a system with two optical feedback loops was studied numerically for a time series prediction task with an information processing rate of 1 GB/s \citep{hou_prediction_2018}. Performance was improved over a single loop configuration owing to the rich dynamics.

\section{Spintronic RC \label{sec:spintronic}}

Several reservoirs based on spin electronics (spintronics) have been proposed. Spintronics is an emerging research field of nanoscale electronics involving both charge and spin of electrons for developing new electronic devices, such as non-volatile storage \citep{wolf2001spintronics}. Spin systems are potential candidates for low-power and small-scale reservoir devices. Three types of spin-based reservoirs with spin oscillations, spin waves, and skyrmions are summarized.

\begin{figure}[t]
\begin{center}
\includegraphics[width=0.8\hsize]{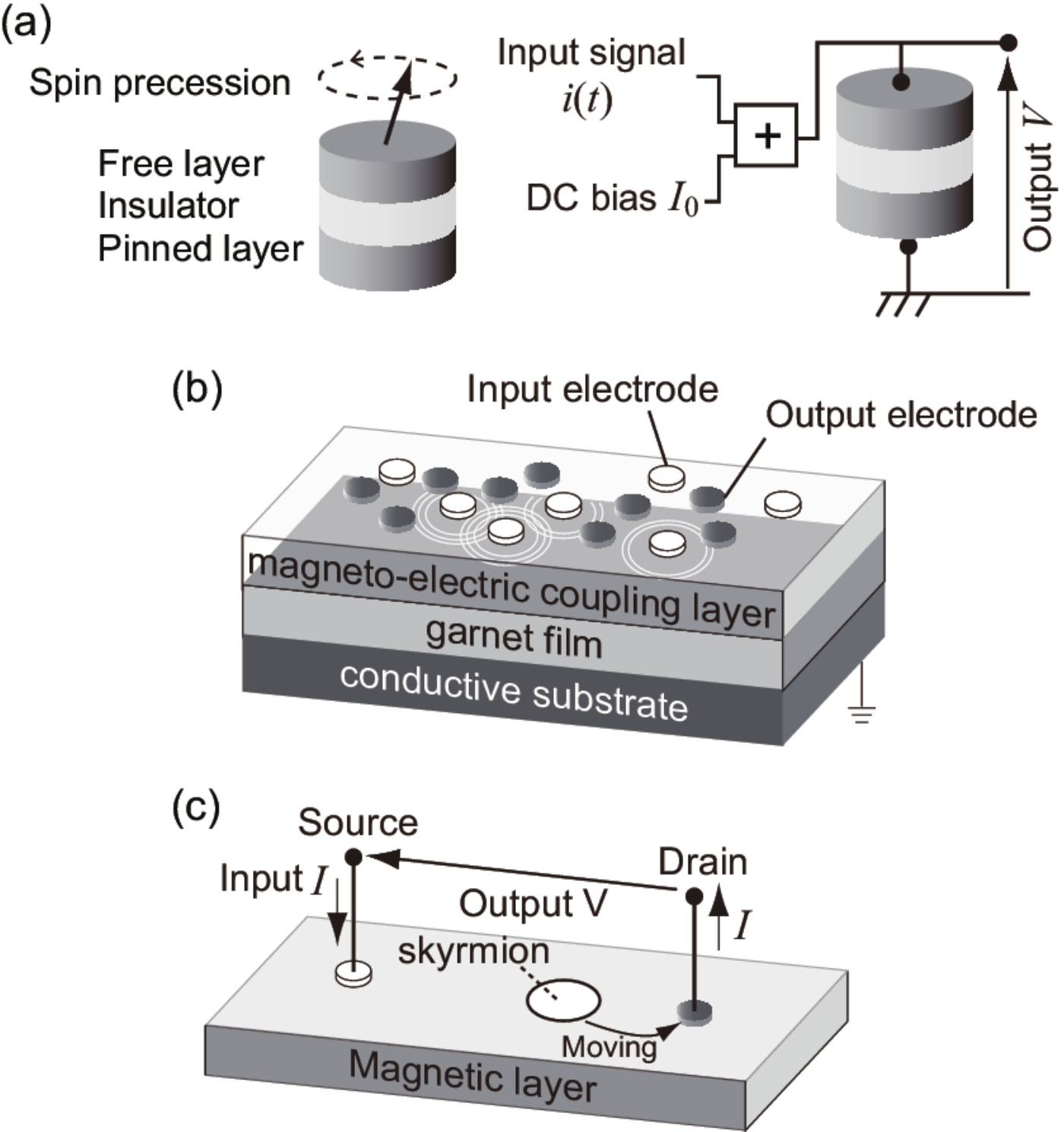}
\caption{(a) Schematic illustration of the spin torque oscillator (left) and the reservoir device structure with a bias current input and a voltage output (right) \citep{torrejon2017neuromorphic}. (b) Schematic illustration of a spin-wave-based reservoir device \citep{nakane2018reservoir}. Figure reproduced with permission from IEEE. (c) A possible structure of a skyrmion-based reservoir device \citep{prychynenko2018magnetic}.} 
\label{fig:spintronic}
\end{center}
\end{figure}

A reservoir with a spin torque oscillator (STO) was experimentally demonstrated \citep{torrejon2017neuromorphic}. The spin torque oscillator is fabricated with a magnetic tunnel junction (MTJ) element composed of two ferromagnets (the top free layer and the bottom pinned layer) separated by a thin insulator, as shown in Fig.~\ref{fig:spintronic}(a). When a constant DC current is injected into an MTJ, the spin direction in the free layer rotates owing to the spin torque that originates from the spin-polarized electron current generated by the pinned layer. After a transient time, the oscillation is stabilized at a frequency depending on the magnitude of the input current. The reservoir uses the nonlinear relationship between the input current and the oscillation frequency as well as history-dependent transient motions of the spin in the free layer. In a spoken digit recognition task, the sound data is given to the single STO as an input current after a preprocessing step and the voltage output of the STO is used to train the readout. It was shown that the digit prediction performance of the proposed method is better than that in the case without STO. The possibility of using chaotic dynamics in an STO with time-delayed feedback for RC has been explored in \citep{williame2017chaotic}. By adjusting the amplitude and delay time of the feedback current, good conditions for spin motions in RC can be specified. Another study numerically investigated the effect of memory and nonlinearity of STO-based reservoirs with single and multiple MTJs in a short-memory task and a parity check task \citep{Furuta2018macromagnetic}. A random binary voltage was used as an input and the time-varying resistance of the MTJ device was used as an output. The spin dynamics in the free layer follows the Landau-Lifshitz-Gilbert (LLG) equation \citep{lakshmanan2011fascinating}. It was demonstrated that the performance in the two tasks depends on the duration of the input pulse voltage.


Spatial propagation of spin motions, or spin waves, can also be used for spintronic reservoir \citep{nakane2018reservoir}. The proposed reservoir device structure consists of a thin Yttrium iron garnet (YIG) film between a magneto-electric coupling layer and a conductive substrate, as shown in Fig.~\ref{fig:spintronic}(b). The input voltage signals applied at the input electrodes on the top layer cause a change in the uniaxial magnetic anisotropy constant of the middle garnet layer due to the magneto-electric effect. This stimulation disturbs the alignment of spins in the same direction; thus, the phase difference between neighboring spins is transmitted through a material as spin waves. The spin motions in a three-dimensional space were simulated by solving the LLG equation. The spin waves can show input history-dependent transient motions and nonlinear interferences \citep{stancil2009spin}. Spin motions measured at the output electrodes were used to train a regression model in the readout. Numerical demonstration showed that a characteristic of the input pattern can be well estimated by the spin-wave-based RC system when the electrode positions and the duration of the spin motions used for computation are appropriately selected.



Another type of spin-based reservoir device could be realized using a magnetic skyrmion, which is a nano-scaled magnetic vortex in a magnetic material, as shown in Fig.~\ref{fig:spintronic}(c) \citep{prychynenko2018magnetic}. It is known that a current-induced transfer of a skyrmion can show nonlinear dynamics and history-dependent spin responses, which are favorable for a reservoir. In the proposed device, the input is an electron current at the source and the output is the voltage between the source and the drain. By injection of a constant current, a skyrmion initially present between the source and the drain moves toward the drain and is then annihilated at the drain. The output voltage changes with the distance between the skyrmion position and the drain, in response to the input current. Interacting multiple skyrmions can generate more complex behavior, and skyrmion fabrics are potentially attractive options for implementing a reservoir \citep{bourianoff2018potential}.

\section{Mechanical RC}\label{sec:mechanical}

Mechanical systems, such as soft and compliant robots, are possible options for physical reservoirs. Soft and compliant robots with flexible bodies are difficult to control due to their complex body dynamics compared with rigid robots with stiff bodies. However, such complex behavior can be favorably leveraged to generate rich nonlinear dynamics required for RC. The idea of outsourcing computation to a physical body is known as {\it morphological computing} in the field of robotics \citep{pfeifer2006body}.



Soft and compliant robots are typically composed of deformable bodies. A primitive model available as a physical reservoir is a mass-spring network, which can be regarded as coupled mechanical oscillators as described in Sec.~\ref{subsec:co}. A mass-spring network reservoir where mass points are randomly connected to neighboring mass points via nonlinear springs was proposed in \citep{hauser2011towards}. The motion of each nonlinear spring can be described as follows:
\begin{eqnarray}
\frac{{\rm d}x_1}{{\rm d}t} &=& x_2, \\
\frac{{\rm d}x_2}{{\rm d}t} &=& -p(x_1) -q(x_2) + u,
\end{eqnarray}
where $t$ represents continuous time, $x_1$ is the displacement of the spring from its rest length, $x_2$ is the velocity of the spring motion, $p$ and $q$ are nonlinear functions representing the properties of the spring, and $u$ is the sum of external forces acting on the spring. The input signal is given to some randomly chosen nodes as the external force, inducing nonlinear responses of the mass-spring network. The output signal is obtained as a linear combination of the actual lengths of the springs. Simulations demonstrated the computing power of RC based on the mass-spring network in time series approximation and robot arm tasks. By adding feedback loops from the output, the reservoir of a mass-spring network can be applied to pattern generation tasks, which are useful for producing locomotion of robots and biological organisms \citep{hauser2012role}. In another extensive study, a reservoir model of a mass-spring-damper network was numerically investigated to clarify the link between the property of the mechanical reservoir and its computational ability in locomotion learning \citep{urbain2017morphological}.

\begin{figure}[t]
\begin{center}
\includegraphics[width=0.9\hsize]{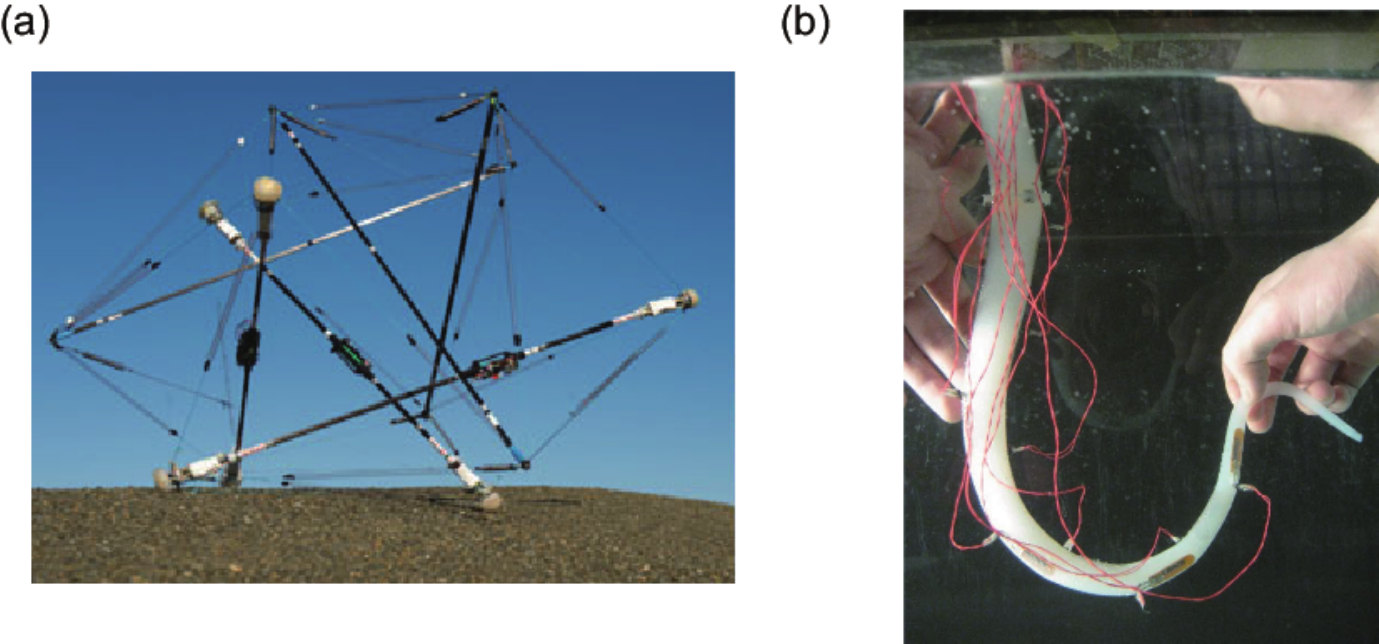}
\caption{(a) A tensegrity structure-based reservoir \citep{caluwaerts2014design}. Figure reproduced with permission from NASA Ames Research Center. (b) A reservoir based on a soft octopus robot \citep{nakajima2015information}. Figure reproduced from \citep{nakajima2015information}, licensed under CC-BY 4.0.}
\label{fig:mechanical}
\end{center}
\end{figure}

By replacing point masses in a mass-spring network by stiff bars, a tensegrity-like structure is obtained. Tensegrity (tension integrity) indicates a structural principle that uses isolated compression elements loaded in a continuous network of tension elements, leading to a physical structure combining strength and flexibility. The compliant body of a tensegrity robot was exploited as a physical reservoir and successfully applied to stable gait pattern generation and terrain pattern classification in numerical experiments \citep{caluwaerts2011body}. Moreover, how to effectively implement control in physical reservoirs with tensegrity structure was investigated by considering various learning rules including a reward-modulated Hebbian learning rule \citep{caluwaerts2013locomotion,burms2015reward}. A physical prototype of a reservoir compliant tensegrity robot (ReCTeR, Fig.~\ref{fig:mechanical}(a)) was developed for creating planetary rovers \citep{caluwaerts2014design}. The highly compliant body of ReCTeR consists of 24 passive spring-cable assemblies and 6 actuated spring-cable assemblies connecting non-parallel struts. The reservoir states obtained from the sensors attached to the struts were used to control the robot in order to approximate desired signals under various control strategies. In another study, better conditions for a tensegrity-based reservoir computer were explored \citep{fujita2018environmental}.

Another type of physical body is a muscular hydrostat system inspired by octopus limbs, as shown in Fig.~\ref{fig:mechanical}(b) \citep{nakajima2013soft}. This is an extreme embodiment of a soft robotic arm without a skeleton, involving virtually unlimited degrees of freedom. Even though the hydrodynamic effect yields highly complex and time-varying dynamics of the soft robotic arm, its motion was successfully learned by an ESN-based controller in a simulation study \citep{kuwabara2012timing} and in an experimental study using a real robot made of silicone rubber \citep{li2012behavior}. These studies suggested that the complex dynamics of the soft body can be used for RC. The computational capability of the soft body was demonstrated in nonlinear system approximations and body dynamics control without an external controller \citep{nakajima2014exploiting,nakajima2015information}.

Other attempts to control real robots and generate different robot behaviors, based on the concept of RC, can be found in studies on a pneumatically driven soft robot arm \citep{eder2017morphological}, a spine-driven quadruped robot \citep{zhao2013spine}, a dog-like quadruped robot \citep{wyffels2010realization}, and a much less compliant quadrupedal robot \citep{degrave2015developing}.
 


\section{Biological RC \label{sec:biological}}
The mechanism behind the computing ability of the brain is one of the mysteries in neuroscience. Considerable efforts have been made to explain the brain computing function using an analogy to computational models in ANNs and machine learning. In attempts to understand the relationship between temporal information processing in the brain and RC, researchers have speculated about which part of the brain can be regarded as a reservoir or a readout as well as about how subnetworks of the brain work in the RC framework. On the other hand, physical RC based on {\it in vitro} biological components has been proposed to investigate the computational capability of biological systems in laboratory experiments. We review recent studies on RC hypotheses in brain regions (Sec.~\ref{subsec:brain}) and RC using {\it in vitro} cultured cells (Sec.~\ref{subsec:vitro}). 

\subsection{Brain regions \label{subsec:brain}}
\begin{figure}[t]
\begin{center}
\includegraphics[width=0.9\hsize]{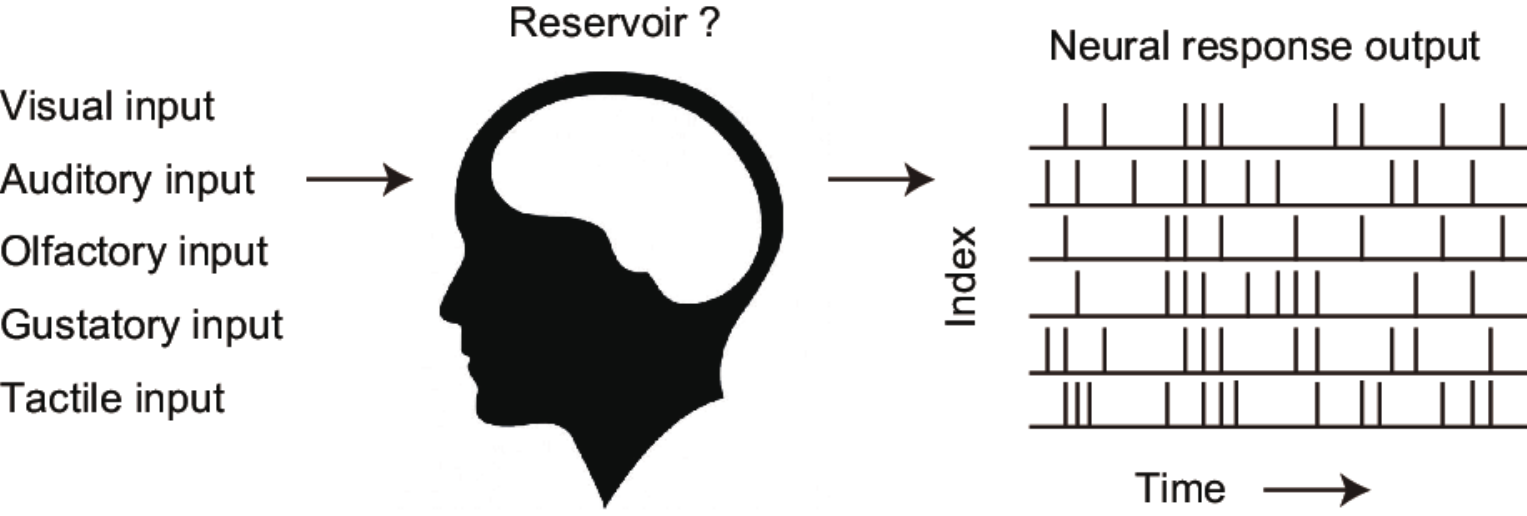}
\caption{A possibility that some brain regions work as reservoirs.}
\label{fig:brain}
\end{center}
\end{figure}

A specific case of RC principle can be found in a series of neurocognitive studies on cortico-striatal models for context-dependent sequential information processing \citep{dominey1995model,dominey1995complex,dominey2013recurrent}. A more general computational framework was later proposed as the LSM, motivated by the real-time information processing of time-varying input streams in cortical microcircuits \citep{maass2002real,maass2004computational} as described in Sec.~\ref{subsec:framework}. Subsequently, researchers have debated on whether reservoir computation is actually conducted in the brain and which brain regions can be interpreted to use this computational framework (Fig.~\ref{fig:brain}). It is suggested that RC is one of the general frameworks for state-dependent computation in cortical networks, emerging from the interaction between the incoming stimulation and the internal dynamics in RNNs \citep{buonomano2009state}. Some researchers have hypothesized that the spatiotemporal information processing in cortical and subcortical networks can be interpreted as RC through neurophysiological experiments and/or computational models, as described below. 

First, we summarize RC-related studies on cortical regions including prefrontal and visual areas.
The prefrontal cortex is a front part of the cerebral cortex, which is associated with the planning of cognitive behavior, personality expression, decision making, and moderating social behavior \citep{fuster2015prefrontal}. An early RC-type model of the cortico-striatal system was proposed to understand the mechanism of context-dependent oculomotor (eye movement) behavior \citep{dominey1995model,dominey1995complex}. The model consists of the prefrontal cortex neurons with fixed recurrent connections (``reservoir'') and modifiable connections from the prefrontal cortex neurons to neurons in the striatum (``readout''). The neural activity in the cortical network, responding to sequential visual inputs, is associated with the outputs that represent the corresponding oculomotor movements through reinforcement learning. An extended work combined the cortico-striatal model (called a temporal recurrent network) for learning serial and temporal structure of sequential inputs with an additional abstract recurrent network which has a short term memory for encoding an abstract structure (or a hidden rule) of sequential inputs \citep{dominey2000neural}. The cortico-striatal model was further combined with neurophysiological models of language processing to learn grammatical constructions in sentence processing \citep{dominey2009cortico,dominey2009neural} and with an RC scheme to improve the performance of the learning algorithm \citep{hinaut2013real}. A detailed history of the cortico-striatal model and its developments is summarized in a review paper \citep{dominey2013recurrent}. On the other hand, the RC properties in monkey prefrontal cortex were investigated with both model simulation and neurophysiological experiments to explore how contexts of sensory inputs are represented in cortical dynamics \citep{enel2016reservoir}. A reservoir of model neurons was compared with the monkey prefrontal cortex in terms of their representational and dynamical properties during a complex cognitive task. Then, it was shown that the reservoir of randomly connected RNNs can obtain a dynamic form of mixed selectivity \citep{rigotti2013importance} and thus perform a complex context-dependent cognitive task as in cortical neurons. This result suggests that RC can be a model of cognitive processing in local generic prefrontal cortical networks. A discussion on the computational role of the thalamus \citep{dehghani2018computational} also highlights the close relationship between the computational properties of the prefrontal cortex and the main features of RC. It is suggested that the context-dependent computing in the prefrontal cortex relies on the task-dependent modulation of the cortical reservoir by thalamic functions.

The visual cortex is a part of the cerebral cortex that processes visual information. The visual cortex located in the occipital lobe is divided into the primary visual cortex (early visual areas) and visual association cortex (higher-order visual areas) \citep{grill2004human}. An experimental study showed evidence that early visual areas have fading memory properties demanded by LSMs \citep{nikolic2009distributed}. In this experiment, {\it in vivo} data of neuronal spiking activities in cat primary visual cortex for different visual stimulus were obtained as a reservoir state and they were classified by a simple linear classifier emulating linear integrate-and-fire readout neurons. The new perspective that the primary visual cortex can perform time-dependent computation for sequential inputs based on memory mechanisms is in contrast to the conventional one that it executes frame-by-frame computation based on memory-less hierarchically organized feedforward architectures \citep{serre2005theory}. The limitation of the conventional viewpoint is that the frame-by-frame computation cannot respond to changes in the context and environment behind visual inputs.  

Second, we move on to studies on RC in subcortical structures including the cerebellum and the basal ganglia. In terms of learning styles, the cerebellum and the basal ganglia can be characterized by supervised learning and reinforcement learning, respectively, which are complementary to each other \citep{doya2000complementary}.
The cerebellum is mainly involved in the coordination of body movement and is responsible for the learning of motor and non-motor skills \citep{rapoport2000role}. The classical hypothesis is that the cerebellum is a learning device, such as a simple perceptron \citep{marr1969theory,albus1971theory}. Another viewpoint from a model-based study is that the functional role of cerebellar circuits is interpreted as an LSM \citep{yamazaki2007cerebellum}. In the proposed model, the granular layer constructed with a recurrent inhibitory circuit is regarded as a reservoir that receives inputs through mossy fibers, and the Purkinje cells are regarded as readout neurons. Numerical studies showed that the cerebellar circuit model successfully learns Boolean functions. This LSM hypothesis highlights an indirect functional pathway from the precerebellar nucleus to the cerebellar nucleus through the LSM model consisting of the granular layer and the Purkinje cells, in addition to the direct pathway from the precerebellar nucleus to the cerebellar nucleus, which is regarded as a simple perceptron as in the classical hypothesis. Further exploration in this direction has made it possible to incorporate recent anatomical and physiological findings on cerebellar microcircuits into computational models \citep{dangelo2016modeling}.

The striatum is the primary input module of the basal ganglia and a critical component of the reward system. It plays a significant role not only in motor control and planning but also in reward-modulated decision making. However, it is not obvious how the environmental states are represented in the reinforcement learning of the basal ganglia. Based on the hypothesis that the striatum responds to diverse inputs from different cortical sources and plays a computational role for discriminating inputs in reinforcement learning-based decision making, the LSM properties of a striatal microcircuit were studied with computational models \citep{toledo2014liquid}. It was demonstrated that the separation and approximation properties required for the LSM are generated using a model network of medium spiny neurons and fast spiking interneurons coupled via inhibitory synapses in a supervised learning task.

Third, we focus on the discussion about the relationship between RC and working memory. Working memory is a cognitive process that temporarily stores and manages information for carrying out complex cognitive tasks \citep{daneman1980individual,baddeley2003working} and is believed to be involved in multiple cortical and subcortical regions \citep{dutta2014neural,eriksson2015neurocognitive}. However, elucidating the mechanism for task-dependent switching of the role of neural circuitry remains an unresolved issue. Multiple working memory models based on RC and their variants have been proposed, including the working memory model with generic cortical microcircuit models with feedback \citep{maass2007computational}, the ESN-based working memory model \citep{pascanu2011neurodynamical}, the reservoir model for storing different time constants of memory traces for reward expectation in reinforcement learning \citep{bernacchia2011reservoir}, the autonomously emerged working memory model through reward-modulated online learning rule that enables the same neural circuit to solve different tasks \citep{hoerzer2012emergence}, and the comparative analysis of three working memory models including the RC-based one \citep{barak2013fixed}.

As described above, the biological plausibility of RC in the brain regions has been examined in many studies, but it requires further investigation. To address this issue, the RC models should be evaluated from multiple aspects of structural, dynamical, and functional properties of cognitive systems. For instance, an investigation focused on the robustness and resilience of reservoir models against structural perturbations in cortical circuits \citep{vincent2016driving}. Examination of RC from the neuroscience viewpoint is expected to be useful for its applications in brain machine interfacing, disease care, and robot control. 



\subsection{{\it in-vitro} cultured cells \label{subsec:vitro}}
A model system to gain insights into the spatiotemporal information processing {\it in vivo} is cultured biological components {\it in vitro}.

\begin{figure}[t]
\begin{center}
\includegraphics[width=0.8\hsize]{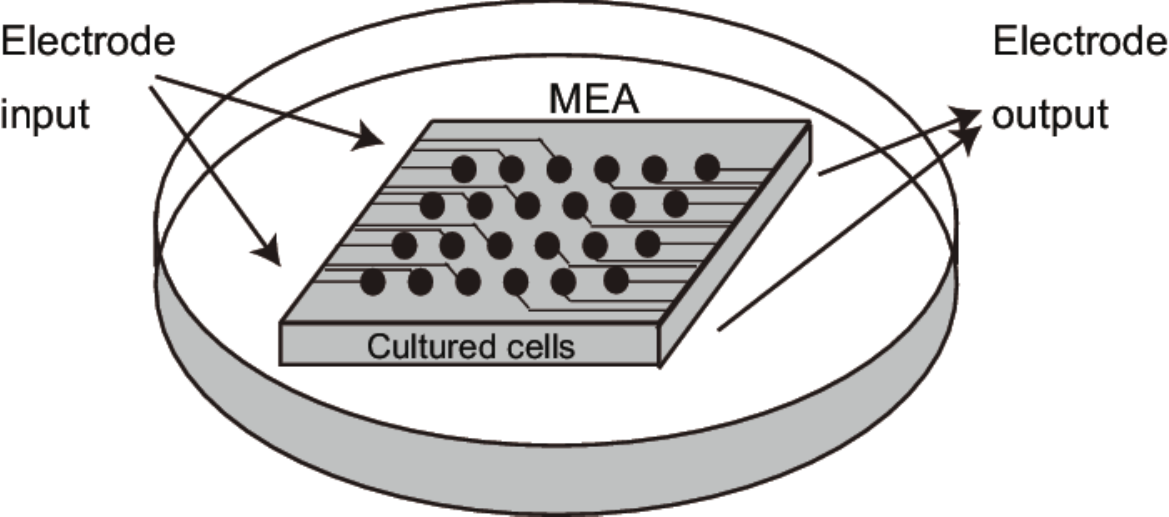}
\caption{A reservoir of {\it in-vitro} cell cultures.}
\label{fig:culture}
\end{center}
\end{figure}

RC with biological neurons has been demonstrated to investigate the computational function of an assembly of biological components. Microelectrode arrays (MEAs) are widely used to electrically stimulate neuronal cultures and measure their responses as shown in Fig.~\ref{fig:culture} \citep{obien2015revealing}. An RC system that combines a reservoir based on cultured cells on MEAs and a computer-aided readout is regarded as a hybrid biological-silicon computer. An early study demonstrated an LSM using {\it in vitro} rat cortical neurons plated on MEAs \citep{hafizovic2007cmos}. Two stimulation patterns with rectangular voltage pulses were given to the electrodes for generating the action potentials of the cultured neurons. The spatiotemporal patterns of spike events recorded with MEAs were transformed into a time-continuous reservoir state by a leaky integrator. In the readout on an external computer, a support vector machine (SVM) was employed for classification of the stimulation patterns. We note that this readout is not a linear classifier and thus the effectiveness of this reservoir remains undetermined. This work was followed by similar experiments. In one such study, the classification ability was investigated with a larger number of stimulation patterns and SVM classifiers, and the tradeoff between the number of patterns and the separability was evaluated \citep{ortman2011input}. In \citep{dockendorf2009liquid}, two types of electrical stimulation protocols were used. One is the low-frequency stimulation inducing bursts spikes and the other is the high-frequency stimulation suppressing the burst response. Living cortical networks react to these stimulus patterns differently, generating distinct spatiotemporal spike trains. The goal is to reconstruct the input spatial stimulation pattern from the temporal structures of the spike trains. 
The results show that the input reconstruction is successful if an appropriate window size is chosen, suggesting the separation property of the biological cell networks. In another study, input coding and output decoding methods for LSMs with cultured neurons have been considered \citep{george2014input}. The proposed input coding method enabled to generate a large number of input patterns from stimulations through a small number of electrodes. The spatiotemporal spike patterns in the reservoir of cultured cells were transformed into spatial patterns such that a linear classifier works well in the readout. 

Instead of electrical stimuli, an optogenetic approach for controlling neuronal activities more precisely has been used in RC experiments. Optical stimuli using random dot patterns were projected onto rat cortical neurons cultured on MEAs as a reservoir \citep{dranias2013short}. Through experiments on short-term memory for optical inputs, it was shown that the results closely match those of corresponding {\it in vivo} experiments in \citep{nikolic2009distributed}. This suggests that the mechanism of state-dependent information processing {\it in vivo} can be studied using cultured neuronal networks. Based on a similar experimental setting, more complex tasks in the time domain were tested \citep{ju2015spatiotemporal}. In classification tasks of spike template patterns and musical styles, high classification performance was achieved if appropriate MEA channels were selected before the readout process and the spatiotemporal memory processes lasted for several seconds.

In addition to the input separability, the control of biological reservoirs has received considerable attention in terms of robot control. In this case, the reservoir receives feedback inputs from the output. The FORCE learning \citep{sussillo2009generating} was used for a closed loop reservoir with cultured neurons to control the output at a target constant value \citep{takahashi2016reservoir}. This technique was adopted to control a mobile robot, by which the robot can avoid obstacles and traverse a maze. An application of the concept of RC to robot control is discussed in \citep{aser2017towards}.

Finally, reservoirs based on other living organisms are discussed. A reservoir with bacterium {\it Escherichia coli} ({\it E.~coli}) reacts to different chemical inputs and generate complex temporal patterns \citep{jones2007there}. A population of {\it E.~coli} was placed in a flask and exposed to sequential inputs that indicate time-varying combinatorial conditions of chemical environments and temperatures. Then, samples of {\it E.~coli} were moved to a microarray to measure the mRNA and protein concentrations, representing the state of the gene regulatory network. Perceptron learning was used in the readout to classify the stimuli into the appropriate classes. Numerical simulations corresponding to the above experimental design confirmed that the proposed RC is partially successful in an XOR problem and shows good performance in a separation problem. This study motivated a corresponding experiment in a wet lab. In \citep{didovyk2014distributed}, genetically engineered bacterial cell cultures were used for realizing a variety of simple classifiers separating different chemical inputs, which are aggregated to form a more complex classifier. This conceptual design of machine learning in synthetic biology can be extended to a reservoir computer.

\section{Others \label{sec:others}}

There are many other candidates for physical reservoirs. Attempts to exploit novel physical reservoirs for RC can contribute to exploration of unconventional computing methods \citep{hadaeghi2017unconventional}. We introduce two types of physical reservoirs that are not categorized into the aforementioned types.

The first one is RC realized with nano-scale materials and substrates that exhibit stimulation-dependent changes. An example is the atomic switch network introduced in Sec.~\ref{subsubsec:mem}. Another proposal is a nano-scale reservoir with quantum dots and chemical compounds that change their absorption spectrum depending on the pH or redox potential in their environment \citep{obst2013nano}. An input signal is given as a change in chemical properties of the compounds, which affect the signal transfer between quantum dots randomly dispersed in a space, encoded as an emission pattern. Simulations confirmed the potential computational performance in an image recognition task. In another study, reservoirs were configured with physical substrates consisting of carbon nanotubes and polymer mixtures \citep{dale2016evolving,dale2016reservoir}. A computer-aided evolutionary algorithm determines the control voltage signals and the locations of input/output electrodes for finding an optimal configuration of the reservoir. Experimental results showed that the material-based RC with the optimized reservoir configuration is successful in time series prediction benchmark tasks.

Another recently proposed idea is to use complex quantum dynamics for RC \citep{fujii2017harnessing}. A quantum system consists of multiple qubits (or quantum bit, indicating the minimum unit of information in quantum computing). For $N$ qubits, there are $2^N$ basis states for a pure quantum state. In quantum RC, each individual node of a quantum reservoir is defined by the system's basis states, not by qubits; therefore, a large number of hidden nodes can be implicitly implemented behind true nodes that are monitored by ensemble measurements. In response to an input signal injected into the first qubit, the quantum system temporally evolves under a Hamiltonian. Then, the states of the qubits obtained by ensemble average measurements are transformed with time multiplexing into a set of signals that are linearly combined to yield an output signal in the readout. High computational performance of the quantum RC was demonstrated in benchmark tasks. In a subsequent study \citep{nakajima2018boosting}, spatial multiplexing was proposed to enhance the computational power of quantum RC by constructing a system of multiple decoupled quantum reservoirs. Physical implementation of a quantum reservoir system is a challenge for the future.

\section{Conclusion and outlook \label{sec:discussion}}

This review has summarized recent trends in RC, with special focus on physical RC. Physical reservoirs have been categorized based on the type of dynamical system (Sec.~\ref{sec:ds}) and physical phenomenon (Secs.~\ref{sec:electronic}-\ref{sec:others}). Most physical reservoirs are designed and configured to satisfy multiple requirements, such as high-dimensionality, nonlinearity, input history-dependent response, fading memory property (echo state property), and/or separation property (Sec.~\ref{subsec:trend}). However, dynamical behavior produced by different types of physical reservoirs exhibits high diversity in terms of the degree of nonlinearity, transient response time, signal transmission speed, and spatial dimension. This diversity makes physical reservoirs appropriate for various tasks and data. 

Physical reservoirs are classified into several types according to the architecture, including network-type reservoir, single-node reservoir with time-delayed feedback, and excitable medium reservoir. They are characterized as follows:
\begin{itemize}
\item Network-type reservoirs consist of interacting nonlinear elements, such as artificial neurons (Sec.~\ref{subsec:framework}), oscillators (Sec.~\ref{subsec:co}), standard circuit elements (Secs.~\ref{subsec:fpga} and \ref{subsec:vlsi}), memristive elements (Sec.~\ref{subsec:memristive}), optical nodes (Sec.~\ref{subsec:opticalnode}), springs (Sec.~\ref{sec:mechanical}), biological cells (Sec.~\ref{subsec:vitro}). They can be simply scaled up by increasing the number of network elements for higher dimensionality. However, realization of a large-scale reservoir requires sophisticated technology to implement massive recurrent interconnections.

\item Single nonlinear node reservoirs with time-delayed feedback generate input-dependent patterns in a high-dimensional space using virtual nodes on a delay loop (Sec.~\ref{subsec:delay}). Such reservoirs have been intensively studied for electronic RC (Secs.~\ref{subsec:analog} and \ref{subsec:fpga}) and photonic RC (Sec.~\ref{subsec:timedelay}). These reservoirs can avoid the problem of massive interconnections and thus are more hardware friendly. However, designing and implementing an appropriate delayed feedback loop is not a straightforward task.

\item Excitable medium reservoirs use propagation of waves triggered by stimulation inputs. Propagation phenomena are widely observed in fluids (Sec.~\ref{subsec:trend}), cellular automata (Sec.~\ref{subsec:ca}), magnetic materials (Sec.~\ref{sec:spintronic}), and elastic media (Sec.~\ref{sec:mechanical}). They have the potential to realize extremely efficient physical RC by harnessing rich physical properties of waves, such as interference, resonance, and synchronization \citep{katayama2016wave}. However, their computational power has not been fully understood thus far. 
\end{itemize}

There are general issues to be addressed in physical RC. First, appropriate preprocessing is necessary for maximizing the computational power of each physical RC system. In fact, some studies have empirically demonstrated that adequate information transformation of input data is necessary for satisfactory performance. In addition, temporal and spatial scaling of input data critically affects the computational performance as the dynamic range is limited in physical reservoirs. Second, each physical reservoir needs to be optimized by the selection of the best-suited material or substrate as well as the tuning of hyper-parameters, such as the shape and size of the reservoir. Mathematical modeling and analysis of a physical reservoir, under physical constraints and practical conditions, are useful for determining the reservoir settings systematically. It is also practically important to ensure scalability in a single reservoir as well as in combined multiple reservoirs. Third, a training algorithm in the readout should be suitably chosen to be compatible with the physical property of RC. Even if a physical reservoir has high signal processing speed, the total computation speed of the entire physical RC system for real-time information processing is limited by the training speed in the readout. One possible solution to this problem is to use a physically realized readout instead of a software-based one. 

Moreover, evaluation of computational performance, processing speed, memory, power efficiency, and scalability of physical RC systems is necessary for comparison with other relevant methods including current digital computers. The computational performance is evaluated with a task-dependent measure, such as classification accuracy in classification tasks and prediction error in prediction tasks. The measures called kernel quality and generalization ability \citep{legenstein2007edge,busing2010connectivity} are also useful for evaluating the effectiveness of a physical reservoir. The processing speed is measured by how much information is processed in a unit time for each task. For instance, in speech recognition tasks, the number of words processed per second under a very low word error rate condition is a measure for processing speed \citep{larger_high-speed_2017}. The memory in reservoirs means the length of time during which the past input information is kept in the reservoir dynamics. The memory capacity is a standard measure for evaluating the memory performance \citep{jaeger2001short}. The power efficiency is evaluated by power consumption required for performing a specific task. For instance, in a study on photonic reservoir \citep{vinckier_high-performance_2015}, the total optical power injected into a passive optical cavity reservoir is considered for assessing its power efficiency aside from power consumption for the other hardware components in preprocessing steps and the readout. The scalability of physical reservoirs is determined by how much the reservoir components can be miniaturized and whether they can be efficiently implemented on a small-scale chip. The evaluation of emerging RC technologies from these different aspects would highlight the advantage and disadvantage of each reservoir. 

Some physical RC systems are suited for realizing next-generation machine learning hardware devices and chips with low power consumption. Such RC hardware is potentially capable of high-speed online computation for dynamic data, in contrast to relatively expensive computation in deep learning hardware for static data. This type of hardware is in great demand as edge computing devices for reducing the communication load in the IoT society, where a massive amount of data is produced and transmitted. Developments of device technology and findings of new materials for this purpose are highly expected. Other physical RC systems are useful for exploring the applicability of natural phenomena to information processing under realistic constraints in biology, chemistry, physics, and engineering. In particular, studies on biological reservoirs are expected to provide new insights into the mechanism of real-time information processing in a variety of brain regions (Sec.~\ref{subsec:brain}).

Physical RC still remains in an early stage of development. It seems premature to compare different physical RC technologies in terms of performance, speed, memory, power efficiency, and scalability, because these characteristics highly depend on the implementation method. It is an ongoing challenge to explore efficient implementation methods for each type of reservoir. For full-scale development of physical RC, further investigation is required from various aspects, such as performance evaluations in practical applications, developments of implementation technology, and theoretical understanding of dynamics and computational function. It is intriguing to exploit novel physical phenomena for RC and to combine physical RC with other machine learning algorithms/hardware. We expect this review to facilitate future interdisciplinary research on physical RC. 

\section*{Acknowledgments}
We would like to thank Ze Hei for his support in collecting the relevant papers and the anonymous reviewers for their useful comments that improved the quality of this review. We would also like to express our gratitude to the action editor for his valuable comments and suggestions. This work was partially supported by JSPS KAKENHI Grant Number JP16H00326 (GT) and partially based on results obtained from a project subsidized by the New Energy and Industrial Technology Development Organization (NEDO).


\end{document}